\documentclass[aps,prl,twocolumn,groupedaddress,showpacs,floatfix,altaffilletter,longbibliography]{revtex4-1}
\usepackage{graphicx}
\usepackage{grffile}
\usepackage{amsmath}
\usepackage{amssymb}
\usepackage{bm}
\usepackage{color}
\usepackage[dvipsnames]{xcolor}
\usepackage{amsmath,amssymb,amsfonts}
\usepackage{epsfig}
\usepackage{times}
\usepackage[colorlinks,bookmarks=false,citecolor=blue,linkcolor=red,urlcolor=blue]{hyperref}
\usepackage{gensymb}
\usepackage[toc,page]{appendix}

\newcommand{\figref}[1]{Fig.~\ref{#1}}
\newcommand{\equref}[1]{Eq.~(\ref{#1})}

\newcommand{\fref}[1]{Fig.~\ref{#1}}
\newcommand{\eref}[1]{Eq.~(\ref{#1})} 

\newcommand{\tref}[1]{Table~\ref{#1}}

\begin{document}

\title{Resilient Fermi liquid and strength of correlations near an antiferromagnetic quantum critical point}

\author{C. Gauvin-Ndiaye$^{1}$, M. Setrakian$^{1}$, and A.-M.S.~Tremblay$^{1}$}
\affiliation{$^1$D{\'e}partement de Physique, Institut quantique, and RQMP Universit{\'e} de Sherbrooke, Sherbrooke, Qu{\'e}bec, Canada  J1K 2R1}
\date{\today}
\begin{abstract}
	Near the antiferromagnetic quantum critical point (QCP) of electron-doped cuprate superconductors, angle-resolved photoemission experiments detect hot spots where the Fermi surface disappears. Here we demonstrate, using the two-particle self-consistent theory, that in the antinodal region the Fermi liquid remains stable for a broad range of angles on the Fermi surface and for all dopings near the QCP. We show how the quasiparticle weight $Z$ and effective mass $m^*$ change and then abruptly become meaningless as the hot spots are approached. We propose a dimensionless number, easily accessible in ARPES experiments, that can be used to gauge the strength of correlations.   
\end{abstract}

\maketitle

\textit{Introduction}.—Fermi liquid theory is the basis on which rests the description of electron behaviour in metals. Although Landau Fermi-liquid theory was formulated for translationally invariant systems, the presence of a lattice leads to relatively small modifications of the original idea. Fermi-liquid quantities, such as the effective mass and the quasiparticle weight, acquire angular dependence along the Fermi surface, a relatively trivial modification. However, the concept of Fermi liquids has been challenged in strongly correlated materials, such as the high-temperature superconductors, where the notions of non-Fermi liquids~\cite{LeeRMP:2006,RoschRMP:2007} and marginal Fermi liquids~\cite{MarginalFermiLiquid:1989} have emerged. The presence of magnetic zero-temperature quantum critical points (QCP) is often invoked as an explanation of non-Fermi liquid behaviour~\cite{RoschRMP:2007}.

Detailed analysis of the Fermi surface quasiparticles through the extraction of the self-energy from ARPES measurements have been performed on multiple materials, such as Sr$_2$RuO$_4$ \cite{Ingle_2005}, organic metals \cite{Kiss_2012}, the hole-doped cuprates Bi$_2$Sr$_2$CaCu$_2$O$_8$ and La$_{2-x}$Sr$_x$CuO$_4$ \cite{Reber_2019,Chang_2013} and the electron-doped cuprate PLCCO \cite{Horio_2020}. In all these cases, Fermi-liquid quasiparticles have been shown to persist in some segments of the Fermi surface, especially away from putative QCPs in {\it overdoped} samples.

Here we show, for the specific case of electron-doped cuprates, where ample experimental data is available~\cite{Armitage_2002,Motoyama_2007,Armitage:2010}, that the proximity to an antiferromagnetic QCP leaves the Fermi liquid unscathed for large portions of the Fermi surface and for all dopings in the vicinity of the QCP. While many theoretical studies have focused on ``hot spots'' where non-Fermi liquid behaviour is observed~\cite{Berg_QMC:2019,Klein_Chubukov_QMC:2020}, we focus on the resilient Fermi-liquid segments that have interesting properties and that can dominate transport~\cite{Sachdev_1995, Hlubina:1995, Lohneysen_2007}. In particular, we show that the properties of the resilient portions of the Fermi liquid lead to a new way to quantify the strength of correlations that goes beyond earlier proposals based, for example, on sum rules~\cite{Basov_Averitt_van_der_Marel_Dressel_Haule_2011}.  

In the electron-doped cuprates, an antiferromagnetic phase extends to high dopings, for example $x\sim0.13$ in Nd$_{2-x}$Ce$_x$CuO$_4$ (NCCO) \cite{Motoyama_2007}. Antiferromagnetic fluctuations play an important role since a loss of spectral weight at the hot spots, connected by the antiferromagnetic (AFM) wave vector, has been observed through ARPES measurements~\cite{Armitage_2002,Matsui_2005,Matsui_2006,He_2019}. Theoretical and experimental proposals have attributed these observations to  antiferromagnetic fluctuations \cite{Kyung_2004,Schafer_2021,Motoyama_2007,Boschini_2020}. Moreover, in the electron-doped cuprate Pr$_{1.3-x}$La$_{0.7}$Ce$_x$CuO$_4$ (PLCCO), it was shown that the suppression of the AFM pseudogap through ``protect annealing'' could be due to the suppression of the AFM fluctuations~\cite{Horio_2016}. 

\textit{Model and Method}.—We study the two-dimensional Hubbard model on a square lattice,
\begin{equation}
H = \sum_{\mathbf{k},\sigma}\epsilon_\mathbf{k} c^{\dagger}_{\mathbf{k}\sigma}c_{\mathbf{k}\sigma} + U\sum_{i}n_{i\uparrow}n_{i\downarrow},
\end{equation}
where $c^{(\dagger)}_{\mathbf{k}\sigma}$ annihilates (creates) an electron of spin $\sigma$ and crystal momentum $\mathbf{k}$. Allowing first, second and third nearest-neighbour hoppings, with respective hopping parameters $t=1$, $t'=-0.175$ and $t''=0.05$ that model the electron-doped cuprate NCCO \cite{Kyung_2004}, the dispersion relation is $\epsilon_\mathbf{k}=-2t (\cos(k_x)+\cos(k_y)) -4t'\cos(k_x)\cos(k_y) -2t''(\cos(2k_x)+\cos(2k_y))$. The strength of interactions is $U$ and $n_{i\uparrow}$,$n_{i\downarrow}$ are number operators for, respectively, spin up and down electrons on site $i$. Planck's constant $\hbar$ is unity. 

We solve the model with the two-particle self-consistent approach (TPSC). This method is non-perturbative and respects conservation laws, the Mermin-Wagner theorem, the Pauli exclusion principle and consistency between single- and two-particle quantities~\cite{Vilk_1997,TremblayMancini:2011}. This method is valid for $U$ ranging from zero to about $0.75$ times the bandwidth. While it cannot reproduce the Mott transition, it enables the study of long-wavelength antiferromagnetic fluctuations and their interactions with electrons. It was the first method to accurately predict the condition under which an AFM pseudogap opens at the hot spots where the AFM Brillouin zone crosses the Fermi surface in the 2D Hubbard model without long-range order. Regions where the Fermi liquid disappears, so-called hot spots, occur when the Vilk criterion is satisfied, namely when the AFM correlation length becomes larger than the thermal de Broglie wavelength \cite{Vilk_1997, Motoyama_2007, Schafer_2021}. A similar phenomenon is seen with the TPSC approach for the attractive Hubbard model, with the prediction of the opening of a pairing pseudogap when the pairing correlation length becomes larger than the de Broglie wavelength \cite{Vilk_1997, Kyung_2001}. In that case, the whole Fermi surface becomes ``hot''. The TPSC approach was previously used to study the electron-doped cuprate NCCO with the band parameters listed above~\cite{Kyung_2004}, reproducing accurately the evolution of the Fermi surface as a function of doping, as observed in ARPES experiments. Unless stated otherwise, we use $U=5.75t$ that was used to reproduce the NCCO ARPES spectra at $x=0.15$.

\begin{figure}
    \centering
    \includegraphics[width=0.95\columnwidth]{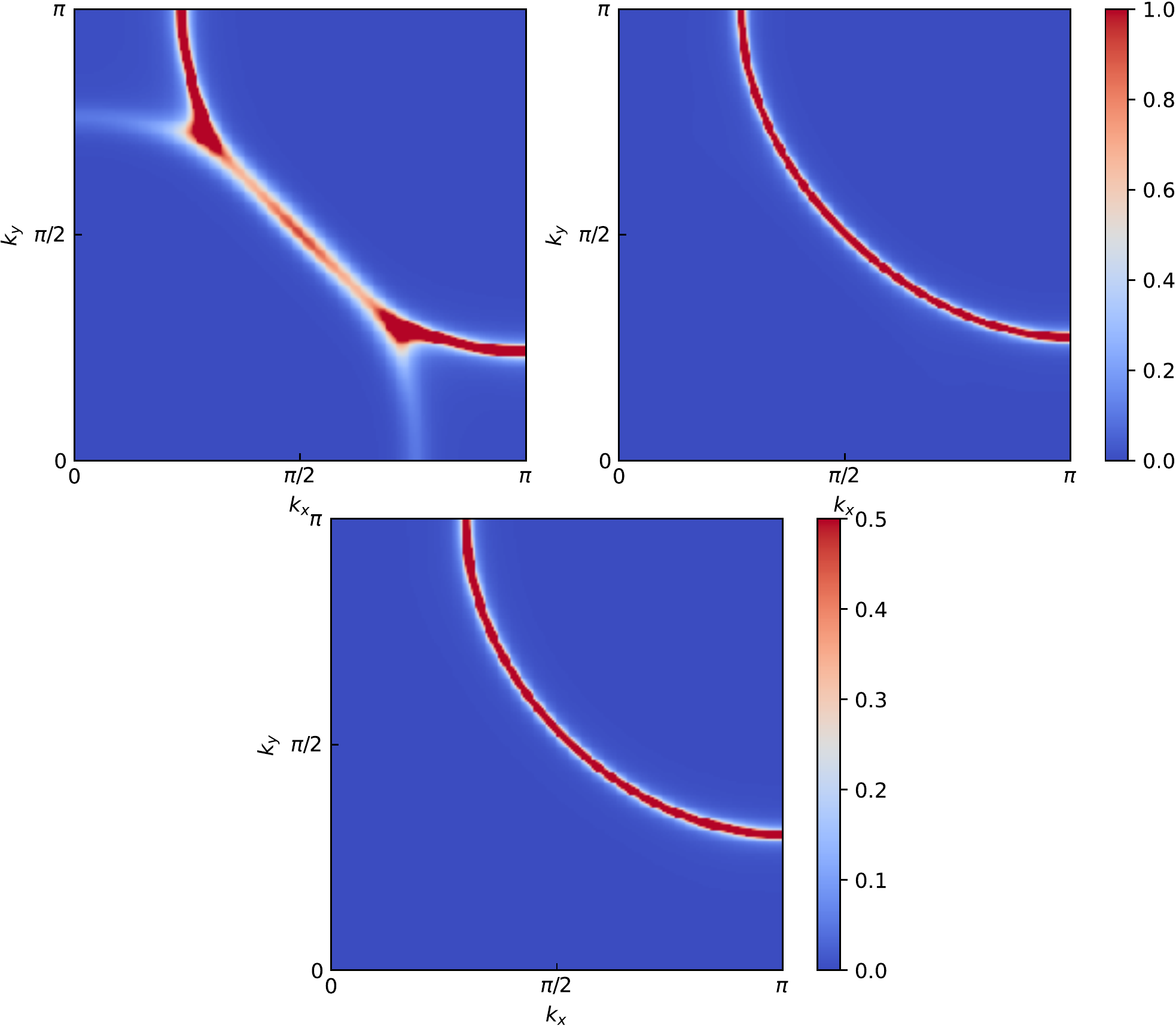} 
    \caption{Momentum distribution curves computed at $T=0.04t$ for $x=0.15$ (top panel, left), $x=0.20$ (top panel, right) and  $x=0.25$ (bottom panel, left).}
    \label{fig:fs}
\end{figure}

\textit{The resilient Fermi liquid}.—To study the effect of AFM fluctuations on Fermi-liquid quasiparticles, we study three different dopings: $x=0.15$ in the underdoped regime (below the AFM QCP), $x=0.20$ near the AFM QCP~\cite{Bergeron_2012}, and $x=0.25$ in the overdoped regime (above the AFM QCP). As shown in \fref{fig:fs}, in the low but non-zero temperature regime, the AFM pseudogap at the hot spots is only present at $x=0.15$, while the Fermi surface, at least when contemplated as a color plot, is well-defined at $x=0.20$ and $x=0.25$. 

\begin{figure}
    \centering
    \includegraphics[width=0.785\columnwidth]{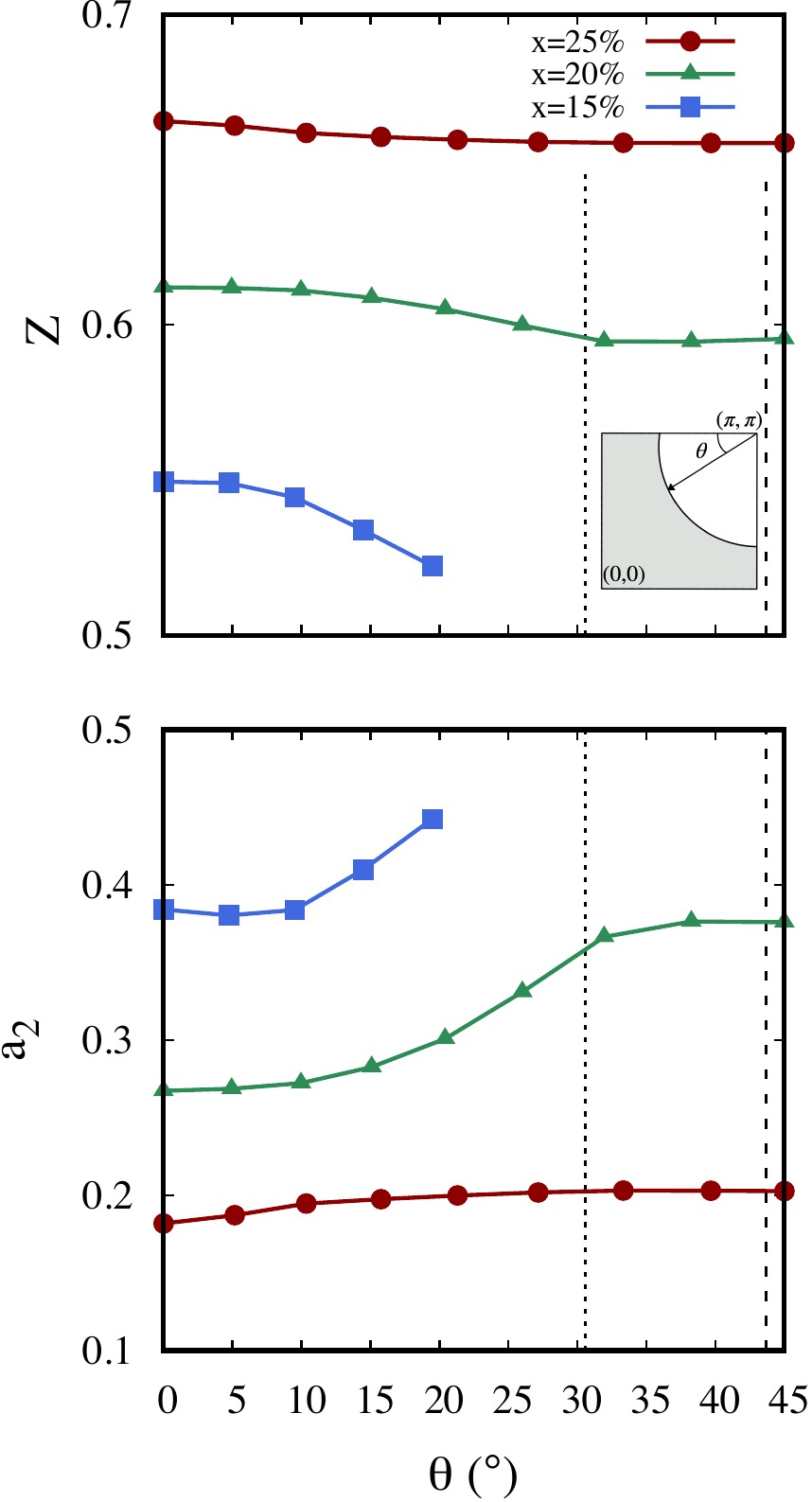} 
    \caption{Fermi-liquid parameters as a function of the angle on the Fermi surface $\theta$ for $x=0.15$ (blue), $x=0.20$ (green) and $x=0.25$ (red). Top panel shows the quasiparticle weight $Z$. Bottom panel shows the $\omega^2$ coefficient $a_2$ of the self-energy in units of $1/t$. The dotted and dashed lines show the hot-spot angle at $x=0.15$ and at $x=0.20$ respectively. The Fermi-surface angle $\theta$ is defined as shown in the insert in the top panel, with $\theta=0$\degree at the antinode and $\theta=45$\degree at the node. The calculations were performed in the temperature range $T=0.02t$ to $T=0.04t$ for $x=0.20$ and $x=0.25$. For $x=0.15$, we did the calculations from $T=0.04t$ to $T=0.05t$ because the TPSC approach is not valid at low temperatures in the renormalized classical regime when the AFM pseudogap becomes too large~\cite{Vilk_1997}. }
    \label{fig:angles}
\end{figure}

We first investigate the behaviour of the quasiparticles as a function of the Fermi-surface angle $\theta$, which is defined in \fref{fig:angles}. In a Fermi liquid, the expected form of the self-energy at low frequency is
\begin{equation}
\Sigma''(\omega,T) = a_0(T) -a_2 \omega^2 ,~\omega < \omega_c 
\label{eq:sigma_reel}
\end{equation}
where $T$ is temperature, $\omega$ is frequency, and $\omega_c$ is a cutoff frequency. The parameters $a_2$, $a_0$ and $\omega_c$ can be extracted from ARPES measurements \cite{Chang_2013, Horio_2020}. In general, the self-energy is momentum-dependent, even though we do not write it explicitly. 

We work in Matsubara frequencies where expansion of the Fermi-liquid self-energy at low frequency gives
\begin{equation}
\Sigma(\omega_n,T) = ia_0(T) + b_0(T) + ia_1 \omega_n + ia_2\omega_n^2 + \mathcal{O}(\omega_n^3),
\label{eq:sigma_mats}
\end{equation}
where $\omega_n = (2n+1)\pi T$ are fermionic Matsubara frequencies and $a_0$, $b_0$, $a_1$ and $a_2$ are real. One can obtain the imaginary part, \eref{eq:sigma_reel}, as well as the real part of the self-energy on the real axis,  
\begin{equation}
    \Sigma'(\omega,T) = b_0(T) + a_1 \omega,
    \label{eq:realsigma_reel}
    \end{equation}
 from the Matsubara expression \eref{eq:sigma_mats} using the analytic continuation $\omega_n \rightarrow - i(\omega+i0^+)$. 

From the real part of the self-energy, \eref{eq:realsigma_reel}, we can extract the quasiparticle weight $Z$
\begin{align}
Z &= \left. \left ( 1 - \frac{\partial \Sigma'(\omega) }{\partial \omega}\right )^{-1} \right |_{\omega=0}, \nonumber\\
&= \frac{1}{1-a_1}. \label{eq:Za1}
\end{align}
The fitting procedure is described in the Supplemental Material~\footnote{See the Supplemental Material at [URL] for the following: Fitting procedure for the Matsubara Self-Energy~\cite{Hodges_1971}; Finding the exponent for the temperature dependence of the scattering rate using a scaling function; The relation between strength of interaction as measured by $U$ and $Z$ and cutoff frequency; Discussion of spectroscopic generalizations of the Kadowaki-Woods ratio~\cite{Jacko_2009}.} \equref{eq:sigma_T} to \equref{eq:SigmaMatsubara}.

\fref{fig:angles} show the quasiparticle weight $Z$ and the coefficient $a_2$ of the Fermi-liquid $\omega^2$ dependence as a function of the Fermi-surface angle $\theta$ for $x=0.15$, $x=0.20$ and $x=0.25$. At the largest doping studied, far from the AFM QCP, both $Z$ and $a_2$ show little dependence on $\theta$. However, we observe a qualitatively different behaviour at $x=0.20$. Even though the Fermi surface is still well-defined at this doping and in the temperature range we study, the effect of antiferromagnetic fluctuations can be seen in the angle dependence of $Z$ and $a_2$. Both parameters have a relatively small dependence in $\theta$ for angles close to the antinode ($0$\degree). However, as $\theta$ increases towards the hot spots and the node, we observe a sharp increase in $a_2$ that is not present in the large doping, $x=0.25$, case. Finally, at $x=0.15$, both $Z$ and $a_2$ are ill-defined for angles near the hot spot. The Fermi-liquid form for the self-energy does not hold near the hot spots and the node when the AFM pseudogap is opened~\cite{Vilk_1997}. Indeed, in this region, the $a_1$ coefficient in $\Sigma'$ cannot be calculated from \eref{eq:sigma_mats} or \equref{eq:a1T} in the Supplemental Material \cite{Note1} since it changes sign near the hot spots, denoting the destruction of the quasiparticles~\cite{Vilk:1996}. However, even with the presence of the AFM pseudogap at $x=0.15$, both $Z$ and $a_2$ remain well-defined near the antinode. The imaginary part of the self-energy at the antinode retains a Fermi-liquid form for this doping, which is situated below the AFM QCP. 

This study of $Z$ and $a_2$ as a function of the Fermi-surface angle points toward an anisotropic destruction of the Fermi-liquid quasiparticles on the Fermi surface, as shown by the survival of well-defined quasiparticles at the antinode below the AFM QCP. Moreover, this destruction seems to be gradual as a function of doping. Before the appearance of the AFM pseudogap, the AFM fluctuations at the QCP already influence the quasiparticles, as illustrated by the increase of $a_2$ near the hot spots at $x=0.20$.


An additional way to determine whether quasiparticles are well-defined or not is to compute the effective mass $m^*$, which can be written as a function of the quasiparticle weight $Z$ and of the momentum dependence of the self-energy perpendicular to the Fermi surface
\begin{align}
\frac{m}{m^*} = Z\left(1+\frac{\partial \Sigma'(\mathbf{k},\omega=0)}{\partial \xi_{\mathbf{k}}}\right ),\\
\frac{\partial \Sigma'(\mathbf{k},\omega=0)}{\partial \xi_{\mathbf{k}}} = \frac{\hat{e}_{\mathbf{k}}\cdot \nabla_{\mathbf{k}}\Sigma'(\mathbf{k},\omega=0)}{\hat{e}_{\mathbf{k}}\cdot \nabla_{\mathbf{k}}\xi_{\mathbf{k}}}.
\label{eq:mass}
\end{align}

We calculate the gradients with small momentum differences $\Delta k$: $\nabla_{\mathbf{k}} \Sigma'(\mathbf{k}) \simeq (\Sigma'(\mathbf{k}\pm\Delta k) - \Sigma'(\mathbf{k}) )/\Delta k$. We assess the ill- or well-defined character of the effective mass by comparing the derivatives calculated with positive and negative $\Delta k$. The results from \eref{eq:mass} for $x=0.15$, $x=0.20$ and $x=0.25$ are shown in \tref{tab:corrections}.
\begin{table}
\begin{center}
    \begin{tabular}{c c  c  c}
    \hline
    \hline
     & $x=0.15$&$x=0.20$&$x=0.25$\\ \hline
         AN, $-\Delta k$ & $2.6$ & $5.2$   & $5.7$  \\
         AN, $+\Delta k$ & $2.8$ & $5.3$ & $5.7$ \\
         N, $-\Delta k$ & $28$ & $15$ & $7.8$\\
         N, $+\Delta k$ & $77$ & $14$ & $7.8$ \\
    \hline \hline
    \end{tabular}
    \caption{ Corrections to the effective mass from the momentum dependence of the self-energy in percentages calculated using \eref{eq:mass}. Calculations were done at $T=0.02$ for $x=0.20$ and $x=0.25$, and at $T=0.04$ for $x=0.15$. The values of $\Delta k$ are $0.12$ for the antinodal direction and $0.17$ for the nodal direction. }\label{tab:corrections}
\end{center}
\end{table}
We observe that the momentum-dependent correction to the effective mass is well-defined at the antinode for all three dopings, since the left and right derivatives are equal and that these corrections are quite small, of the order of $5$\%. This is also true at the node at $x=0.25$. At the node at $x=0.20$, however, the correction is larger by an order of magnitude. 
This indicates a strong momentum dependence for this doping at the node, even without an AFM pseudogap. At the node at $x=0.15$, the correction is large and the left and right derivatives disagree. These findings are consistent with our previous discussion of $Z$ and $a_2$, and reinforce our conclusion that Fermi-liquid quasiparticles remain well-defined at the antinode at $x=0.15$.

\textit{Temperature dependence of the scattering rate}.—Further signatures of the anisotropy between the node and the antinode at the QCP, $x=0.20$, can be found through the temperature scaling of the self-energy. Following appendix B of Ref~\cite{Schafer_2021}, we assume $\omega/T$ scaling for the imaginary part of the self-energy in real frequencies
\begin{equation}
\Sigma''(\omega) = a(\pi T)^{\nu}\phi\left(\frac{\omega}{\pi T}\right),
\label{eq:scaling}
\end{equation}
where $a$ is a constant. This allows us to find the exponent $\nu$ at $\omega=0$ with the procedure described in the Supplemental Material~\cite{Note1}, Eq. (S6) to Eq. (S11). The exponent $\nu=2$ corresponds to Fermi-liquid behaviour, while $\nu<2$ corresponds to non-Fermi liquid behaviour. 

We calculate the exponent $\nu$ at the node and at the antinode for $x=0.20$ and $x=0.25$. At the antinode, we recover $\nu \simeq 2$ for both dopings. While $\nu \simeq 2$ is also true at the node for $x=0.25$, it is not the case at $x=0.20$, where $\nu \simeq 1.4$. Similarly to our calculation of $a_2$ as a function of the Fermi-surface angle, this scaling analysis shows that the Fermi liquid is resilient only near the antinode at the AFM QCP ($x=0.20$).  


\textit{Fermi liquid and strength of interactions}.—The properties of the antinodal Fermi liquid can be used to quantify the strength of interactions. It has been proposed that the Fermi-liquid cutoff frequency $\omega_c$, namely the frequency at which $\omega^2$ behaviour disappears, can indicate the strength of the interactions in a correlated material \cite{Horio_2020}. We focus on the antinodal direction for $x=0.15$, $x=0.20$ and $x=0.25$, and on the nodal direction for $x=0.25$, where the Fermi liquid is stable. We vary $U$ from $1t$ to $5t$. Evidently, the determination of the cutoff frequency $\omega_c$ has some arbitrariness, but as long as one adheres to a definition, the results are consistent. Here, we calculate the deviation between the imaginary part of the linearly-interpolated Matsubara self-energy we obtain from our calculation and the fit using \eref{eq:sigma_mats}. We take $\omega_c$ as the frequency at which this deviation reaches a threshold of $10\%$, $15\%$ or $20$\% for a fixed temperature $T=0.02$. 

As shown in \figref{fig:U} of the Supplemental Material~\cite{Note1}, $\omega_c$ decreases with $U$ and increases with the quasiparticle weight $Z$, so that $\omega_c$ can indeed measure the strength of interactions. Moreover, for a fixed value of $U$, $\omega_c$ increases with doping, which means that $\omega_c$ is a measure of electronic correlations in a broader sense. Given the angular dependence of $Z$ illustrated in \fref{fig:angles}, it is clear that this measure of interaction is not uniquely defined for a given compound.      

\begin{figure}[h]
    \centering
    \includegraphics[width=1.0\columnwidth]{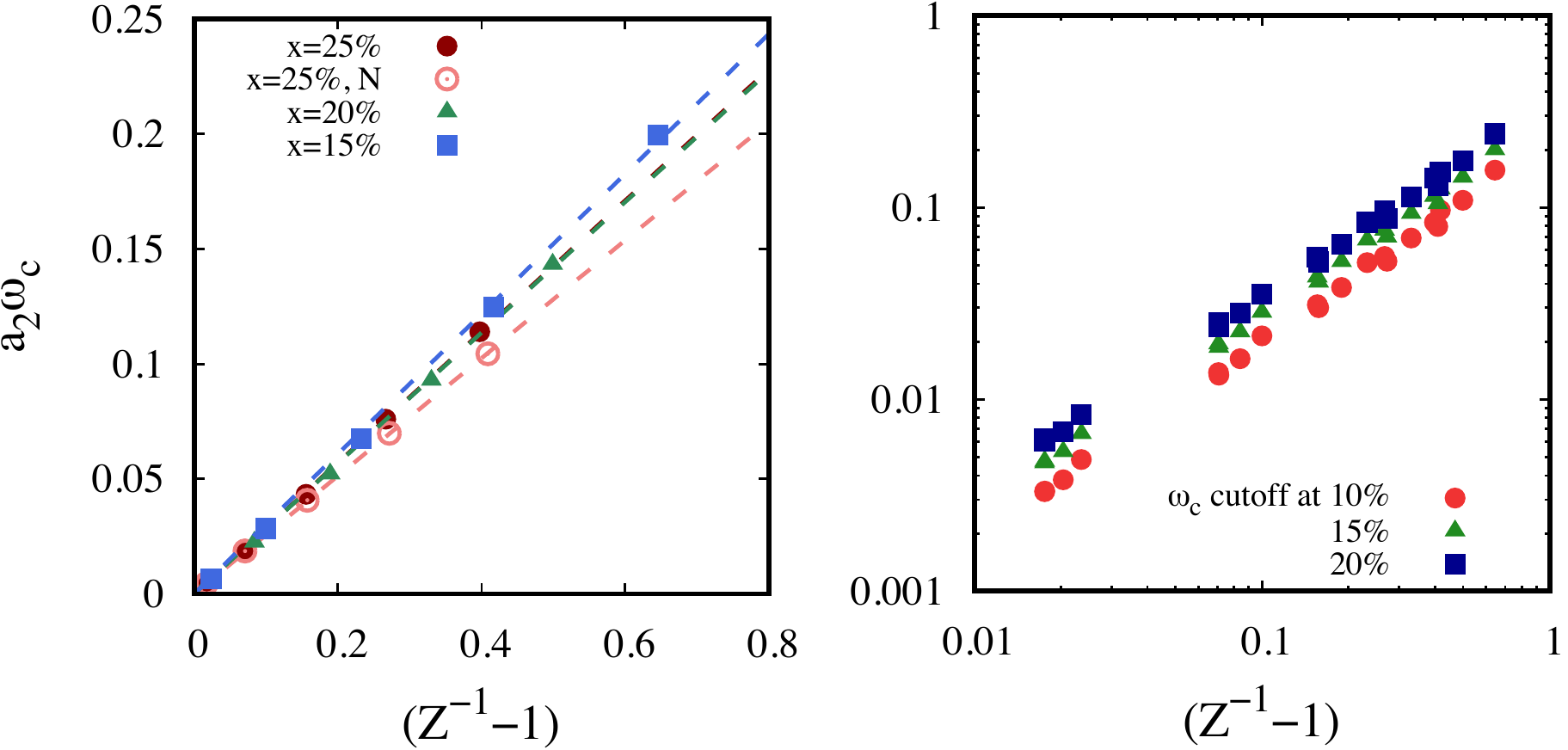} 
    \caption{ 
    The product of $a_2$, the coefficient of $\omega^2$ in $\Sigma''$, and of $\omega_c$, the Fermi-liquid cutoff frequency, is related to the quasiparticle weight $Z$ through a Kramers-Kronig relation. Here, all the dots are computed for different values of $U$ (shown in \fref{fig:U}). Left panel shows data at the antinode, for dopings $x=0.25$ (red), $x=0.20$ (green) and $x=0.15$ (blue), and at the node for $x=0.25$ (pink), with $\omega_c$ calculated with a relative deviation threshold from $\omega_n^2$ of $15$\%. The right panel shows a log-log plot of the data for the dopings listed above, with $\omega_c$ calculated with three different thresholds, $10$\% (orange), $15$\% (green) and $20$\% (blue). Note that for a given threshold, all dopings of the left panel here fall on the same straight line, with only small variations. 
    }
    \label{fig:a2vsZm1}
\end{figure}

Horio {\it et al.} \cite{Horio_2020} also suggested a spectroscopic analog of the Kadowaki-Woods ratio that would relate $a_2$ and $Z$, two quantities we obtained in our calculations for multiple values of $U$. This is discussed in \fref{fig:KW}. We propose instead that the dimensionless number $a_2\times\omega_c$ is a more robust estimate that scales as $(Z^{-1}-1)$. It is shown in \figref{fig:a2vsZm1} (left panel) with $\omega_c$ calculated using a threshold of $15$\%. This scaling is a general result that follows from the Kramers-Kronig relation. More precisely, we find

\begin{equation}
    Z_{\mathbf{k}}^{-1}-1 = \frac{4 \xi_{\mathbf{k}}}{\pi}\omega_{c,\mathbf{k}} a_{2,\mathbf{k}},
    \label{eq:a2vsZm1}
\end{equation}
where $\xi_{\mathbf{k}}$ is a constant that varies between $2$ and $4$ in our calculations. It can be both momentum- and material-dependent, as seen from the different slopes in the left panel of \fref{fig:a2vsZm1}. It also depends on the criterion that defines the cutoff frequency $\omega_c$. This highlights the challenge in finding $Z$ from spectroscopic data. Nevertheless, for general trends, it is quite useful as shown on the right panel of \fref{fig:a2vsZm1}. There we show on a log-log plot the scaling $a_2\times\omega_c$ \textit{vs} $(Z^{-1}-1)$ for the different dopings and angles as a function of the threshold for the cutoff frequency $\omega_c$. On this scale, only the choice of threshold for the cutoff frequency, shown in different colors, has a visible effect on the result. 

\textit{Discussion}.—One could argue that the resilience of the antinodal Fermi liquid is not surprising given that the electron Fermi-surface pocket of the $T=0$ antiferromagnet basically coincides for a large part with the antinodal section in the normal state. This is not so trivial, however, since a hole pocket also develops in the nodal direction in the AFM that eventually occurs at $T=0$, while that part of the Fermi surface disappears completely in the pseudogap regime, at least for the temperatures we could consider.

An angle dependence of the $a_2$ coefficient analogous to what we observe here has been previously measured in the hole-doped cuprate LSCO in the overdoped regime, outside of the pseudogap phase \cite{Chang_2013}. In the case of LSCO, the $a_2$ coefficient was found to be stable around the node and to increase as the angle decreases toward the antinode, before vanishing at an angle of $\phi_0\simeq 15$\degree. This behaviour is analogous to the one described here, but the roles of the antinode and the node are exchanged. This is expected from the fact that, in the hole-doped cuprates, the pseudogap opens in the antinodal region of the Fermi surface. 

In contrast, recent ARPES experiments have shown that the $a_2$ coefficient in the electron-doped cuprate PLCCO is constant as a function of the Fermi-surface angle in the overdoped regime \cite{Horio_2020}. These measurements are reminiscent of our results at $x=0.25$, far from the AFM QCP. The case of LSCO also illustrates that different portions of the Fermi surface can be affected very differently by interactions, analogous to what we have seen in our calculations (\fref{fig:angles}).    

 Our plot \fref{fig:a2vsZm1}, based on the stable Fermi-liquid portions of the various compounds, quantifies the relative strength of interactions. The two quantities $a_2$, the coefficient of the $\omega^2$ dependence, and $\omega_c$ the cutoff frequency for Fermi-liquid behaviour, can be obtained experimentally from ARPES data. Given a future agreement between researchers on a reference case and on the way the cutoff frequency $\omega_c$ is determined, it becomes possible to compare the correlation strength in different materials using the proportionality between $a_2 \times \omega_c$ and $Z^{-1}-1$. At this stage, we cannot compare quantitatively our results to experimental measurements because we obtained $\omega_c$ from the Matsubara self-energy. Nevertheless, using the data presented in Refs. \cite{Horio_2020, Chang_2013}, we obtain $a_2\times\omega_c = 0.90\pm0.07$ for PLCCO while for LSCO, depending on the Fermi-surface angle, we obtain $a_2\times\omega_c$ ranging from $1.0\pm0.4$ to $1.7\pm0.2$. This supports previous theoretical suggestions that hole-doped cuprates are more strongly correlated than electron-doped cuprates \cite{Senechal_2004, Hankevych_2006, Weber_2010}. 

\textit{Conclusion}.—The stability of the Fermi liquid on portions of the Fermi surface should be a general property of materials where electrons scatter off critical fluctuations with non-zero wave vector~\cite{Hlubina:1995}. In such a case, the hot-spot phenomenon occurs and ``cold regions'' are bound to be stable Fermi liquids whose product $a_2 \times \omega_c$, accessible in ARPES experiments, can be used to provide a dimensionless scale that quantifies the strength of interactions. 

\paragraph{Acknowledgments} We are grateful to P.-A. Graham, M. Horio and J. Chang for useful discussions and for sharing their data. This work has been supported by the Natural Sciences and Engineering Research Council of Canada (NSERC) under grant RGPIN-2019-05312, by a Vanier Scholarship (C. G.-N.) from NSERC and by the Canada First Research Excellence Fund. Simulations were performed on computers provided by the Canadian Foundation for Innovation, the Minist\`ere de l'\'Education des Loisirs et du Sport (Qu\'ebec), Calcul Qu\'ebec, and Compute Canada.


\begin{thebibliography}{35}%
    \makeatletter
    \providecommand \@ifxundefined [1]{%
     \@ifx{#1\undefined}
    }%
    \providecommand \@ifnum [1]{%
     \ifnum #1\expandafter \@firstoftwo
     \else \expandafter \@secondoftwo
     \fi
    }%
    \providecommand \@ifx [1]{%
     \ifx #1\expandafter \@firstoftwo
     \else \expandafter \@secondoftwo
     \fi
    }%
    \providecommand \natexlab [1]{#1}%
    \providecommand \enquote  [1]{``#1''}%
    \providecommand \bibnamefont  [1]{#1}%
    \providecommand \bibfnamefont [1]{#1}%
    \providecommand \citenamefont [1]{#1}%
    \providecommand \href@noop [0]{\@secondoftwo}%
    \providecommand \href [0]{\begingroup \@sanitize@url \@href}%
    \providecommand \@href[1]{\@@startlink{#1}\@@href}%
    \providecommand \@@href[1]{\endgroup#1\@@endlink}%
    \providecommand \@sanitize@url [0]{\catcode `\\12\catcode `\$12\catcode
      `\&12\catcode `\#12\catcode `\^12\catcode `\_12\catcode `\%12\relax}%
    \providecommand \@@startlink[1]{}%
    \providecommand \@@endlink[0]{}%
    \providecommand \url  [0]{\begingroup\@sanitize@url \@url }%
    \providecommand \@url [1]{\endgroup\@href {#1}{\urlprefix }}%
    \providecommand \urlprefix  [0]{URL }%
    \providecommand \Eprint [0]{\href }%
    \providecommand \doibase [0]{http://dx.doi.org/}%
    \providecommand \selectlanguage [0]{\@gobble}%
    \providecommand \bibinfo  [0]{\@secondoftwo}%
    \providecommand \bibfield  [0]{\@secondoftwo}%
    \providecommand \translation [1]{[#1]}%
    \providecommand \BibitemOpen [0]{}%
    \providecommand \bibitemStop [0]{}%
    \providecommand \bibitemNoStop [0]{.\EOS\space}%
    \providecommand \EOS [0]{\spacefactor3000\relax}%
    \providecommand \BibitemShut  [1]{\csname bibitem#1\endcsname}%
    \let\auto@bib@innerbib\@empty
    \bibitem [{\citenamefont {Lee}\ \emph {et~al.}(2006)\citenamefont {Lee},
      \citenamefont {Nagaosa},\ and\ \citenamefont {Wen}}]{LeeRMP:2006}%
      \BibitemOpen
      \bibfield  {author} {\bibinfo {author} {\bibfnamefont {Patrick~A.}\
      \bibnamefont {Lee}}, \bibinfo {author} {\bibfnamefont {Naoto}\ \bibnamefont
      {Nagaosa}}, \ and\ \bibinfo {author} {\bibfnamefont {Xiao-Gang}\ \bibnamefont
      {Wen}},\ }\bibfield  {title} {\enquote {\bibinfo {title} {Doping a mott
      insulator: Physics of high-temperature superconductivity},}\ }\href {\doibase
      10.1103/RevModPhys.78.17} {\bibfield  {journal} {\bibinfo  {journal} {Rev.
      Mod. Phys.}\ }\textbf {\bibinfo {volume} {78}},\ \bibinfo {pages} {17--85}
      (\bibinfo {year} {2006})}\BibitemShut {NoStop}%
    \bibitem [{\citenamefont {L\"ohneysen}\ \emph
      {et~al.}(2007{\natexlab{a}})\citenamefont {L\"ohneysen}, \citenamefont
      {Rosch}, \citenamefont {Vojta},\ and\ \citenamefont
      {W\"olfle}}]{RoschRMP:2007}%
      \BibitemOpen
      \bibfield  {author} {\bibinfo {author} {\bibfnamefont {Hilbert~v.}\
      \bibnamefont {L\"ohneysen}}, \bibinfo {author} {\bibfnamefont {Achim}\
      \bibnamefont {Rosch}}, \bibinfo {author} {\bibfnamefont {Matthias}\
      \bibnamefont {Vojta}}, \ and\ \bibinfo {author} {\bibfnamefont {Peter}\
      \bibnamefont {W\"olfle}},\ }\bibfield  {title} {\enquote {\bibinfo {title}
      {Fermi-liquid instabilities at magnetic quantum phase transitions},}\ }\href
      {\doibase 10.1103/RevModPhys.79.1015} {\bibfield  {journal} {\bibinfo
      {journal} {Rev. Mod. Phys.}\ }\textbf {\bibinfo {volume} {79}},\ \bibinfo
      {pages} {1015--1075} (\bibinfo {year} {2007}{\natexlab{a}})}\BibitemShut
      {NoStop}%
    \bibitem [{\citenamefont {Varma}\ \emph {et~al.}(1989)\citenamefont {Varma},
      \citenamefont {Littlewood}, \citenamefont {Schmitt-Rink}, \citenamefont
      {Abrahams},\ and\ \citenamefont {Ruckenstein}}]{MarginalFermiLiquid:1989}%
      \BibitemOpen
      \bibfield  {author} {\bibinfo {author} {\bibfnamefont {C.~M.}\ \bibnamefont
      {Varma}}, \bibinfo {author} {\bibfnamefont {P.~B.}\ \bibnamefont
      {Littlewood}}, \bibinfo {author} {\bibfnamefont {S.}~\bibnamefont
      {Schmitt-Rink}}, \bibinfo {author} {\bibfnamefont {E.}~\bibnamefont
      {Abrahams}}, \ and\ \bibinfo {author} {\bibfnamefont {A.~E.}\ \bibnamefont
      {Ruckenstein}},\ }\bibfield  {title} {\enquote {\bibinfo {title}
      {Phenomenology of the normal state of cu-o high-temperature
      superconductors},}\ }\href {\doibase 10.1103/PhysRevLett.63.1996} {\bibfield
      {journal} {\bibinfo  {journal} {Phys. Rev. Lett.}\ }\textbf {\bibinfo
      {volume} {63}},\ \bibinfo {pages} {1996--1999} (\bibinfo {year}
      {1989})}\BibitemShut {NoStop}%
    \bibitem [{\citenamefont {Ingle}\ \emph {et~al.}(2005)\citenamefont {Ingle},
      \citenamefont {Shen}, \citenamefont {Baumberger}, \citenamefont {Meevasana},
      \citenamefont {Lu}, \citenamefont {Shen}, \citenamefont {Damascelli},
      \citenamefont {Nakatsuji}, \citenamefont {Mao}, \citenamefont {Maeno},\ and\
      \citenamefont {et~al.}}]{Ingle_2005}%
      \BibitemOpen
      \bibfield  {author} {\bibinfo {author} {\bibfnamefont {N.~J.~C.}\
      \bibnamefont {Ingle}}, \bibinfo {author} {\bibfnamefont {K.~M.}\ \bibnamefont
      {Shen}}, \bibinfo {author} {\bibfnamefont {F.}~\bibnamefont {Baumberger}},
      \bibinfo {author} {\bibfnamefont {W.}~\bibnamefont {Meevasana}}, \bibinfo
      {author} {\bibfnamefont {D.~H.}\ \bibnamefont {Lu}}, \bibinfo {author}
      {\bibfnamefont {Z.-X.}\ \bibnamefont {Shen}}, \bibinfo {author}
      {\bibfnamefont {A.}~\bibnamefont {Damascelli}}, \bibinfo {author}
      {\bibfnamefont {S.}~\bibnamefont {Nakatsuji}}, \bibinfo {author}
      {\bibfnamefont {Z.~Q.}\ \bibnamefont {Mao}}, \bibinfo {author} {\bibfnamefont
      {Y.}~\bibnamefont {Maeno}}, \ and\ \bibinfo {author} {\bibnamefont
      {et~al.}},\ }\bibfield  {title} {\enquote {\bibinfo {title} {Quantitative
      analysis of sr 2 ruo 4 angle-resolved photoemission spectra: Many-body
      interactions in a model fermi liquid},}\ }\href {\doibase
      10.1103/PhysRevB.72.205114} {\bibfield  {journal} {\bibinfo  {journal}
      {Physical Review B}\ }\textbf {\bibinfo {volume} {72}},\ \bibinfo {pages}
      {205114} (\bibinfo {year} {2005})}\BibitemShut {NoStop}%
    \bibitem [{\citenamefont {Kiss}\ \emph {et~al.}(2012)\citenamefont {Kiss},
      \citenamefont {Chainani}, \citenamefont {Yamamoto}, \citenamefont {Miyazaki},
      \citenamefont {Akimoto}, \citenamefont {Shimojima}, \citenamefont {Ishizaka},
      \citenamefont {Watanabe}, \citenamefont {Chen}, \citenamefont {Fukaya},\ and\
      \citenamefont {et~al.}}]{Kiss_2012}%
      \BibitemOpen
      \bibfield  {author} {\bibinfo {author} {\bibfnamefont {T.}~\bibnamefont
      {Kiss}}, \bibinfo {author} {\bibfnamefont {A.}~\bibnamefont {Chainani}},
      \bibinfo {author} {\bibfnamefont {H.M.}\ \bibnamefont {Yamamoto}}, \bibinfo
      {author} {\bibfnamefont {T.}~\bibnamefont {Miyazaki}}, \bibinfo {author}
      {\bibfnamefont {T.}~\bibnamefont {Akimoto}}, \bibinfo {author} {\bibfnamefont
      {T.}~\bibnamefont {Shimojima}}, \bibinfo {author} {\bibfnamefont
      {K.}~\bibnamefont {Ishizaka}}, \bibinfo {author} {\bibfnamefont
      {S.}~\bibnamefont {Watanabe}}, \bibinfo {author} {\bibfnamefont {C.-T.}\
      \bibnamefont {Chen}}, \bibinfo {author} {\bibfnamefont {A.}~\bibnamefont
      {Fukaya}}, \ and\ \bibinfo {author} {\bibnamefont {et~al.}},\ }\bibfield
      {title} {\enquote {\bibinfo {title} {Quasiparticles and fermi liquid
      behaviour in an organic metal},}\ }\href {\doibase 10.1038/ncomms2079}
      {\bibfield  {journal} {\bibinfo  {journal} {Nature Communications}\ }\textbf
      {\bibinfo {volume} {3}},\ \bibinfo {pages} {1089} (\bibinfo {year}
      {2012})}\BibitemShut {NoStop}%
    \bibitem [{\citenamefont {Reber}\ \emph {et~al.}(2019)\citenamefont {Reber},
      \citenamefont {Zhou}, \citenamefont {Plumb}, \citenamefont {Parham},
      \citenamefont {Waugh}, \citenamefont {Cao}, \citenamefont {Sun},
      \citenamefont {Li}, \citenamefont {Wang}, \citenamefont {Wen},\ and\
      \citenamefont {et~al.}}]{Reber_2019}%
      \BibitemOpen
      \bibfield  {author} {\bibinfo {author} {\bibfnamefont {T.~J.}\ \bibnamefont
      {Reber}}, \bibinfo {author} {\bibfnamefont {X.}~\bibnamefont {Zhou}},
      \bibinfo {author} {\bibfnamefont {N.~C.}\ \bibnamefont {Plumb}}, \bibinfo
      {author} {\bibfnamefont {S.}~\bibnamefont {Parham}}, \bibinfo {author}
      {\bibfnamefont {J.~A.}\ \bibnamefont {Waugh}}, \bibinfo {author}
      {\bibfnamefont {Y.}~\bibnamefont {Cao}}, \bibinfo {author} {\bibfnamefont
      {Z.}~\bibnamefont {Sun}}, \bibinfo {author} {\bibfnamefont {H.}~\bibnamefont
      {Li}}, \bibinfo {author} {\bibfnamefont {Q.}~\bibnamefont {Wang}}, \bibinfo
      {author} {\bibfnamefont {J.~S.}\ \bibnamefont {Wen}}, \ and\ \bibinfo
      {author} {\bibnamefont {et~al.}},\ }\bibfield  {title} {\enquote {\bibinfo
      {title} {A unified form of low-energy nodal electronic interactions in
      hole-doped cuprate superconductors},}\ }\href {\doibase
      10.1038/s41467-019-13497-4} {\bibfield  {journal} {\bibinfo  {journal}
      {Nature Communications}\ }\textbf {\bibinfo {volume} {10}},\ \bibinfo {pages}
      {5737} (\bibinfo {year} {2019})}\BibitemShut {NoStop}%
    \bibitem [{\citenamefont {Chang}\ \emph {et~al.}(2013)\citenamefont {Chang},
      \citenamefont {Månsson}, \citenamefont {Pailhès}, \citenamefont {Claesson},
      \citenamefont {Lipscombe}, \citenamefont {Hayden}, \citenamefont {Patthey},
      \citenamefont {Tjernberg},\ and\ \citenamefont {Mesot}}]{Chang_2013}%
      \BibitemOpen
      \bibfield  {author} {\bibinfo {author} {\bibfnamefont {J.}~\bibnamefont
      {Chang}}, \bibinfo {author} {\bibfnamefont {M.}~\bibnamefont {Månsson}},
      \bibinfo {author} {\bibfnamefont {S.}~\bibnamefont {Pailhès}}, \bibinfo
      {author} {\bibfnamefont {T.}~\bibnamefont {Claesson}}, \bibinfo {author}
      {\bibfnamefont {O.~J.}\ \bibnamefont {Lipscombe}}, \bibinfo {author}
      {\bibfnamefont {S.~M.}\ \bibnamefont {Hayden}}, \bibinfo {author}
      {\bibfnamefont {L.}~\bibnamefont {Patthey}}, \bibinfo {author} {\bibfnamefont
      {O.}~\bibnamefont {Tjernberg}}, \ and\ \bibinfo {author} {\bibfnamefont
      {J.}~\bibnamefont {Mesot}},\ }\bibfield  {title} {\enquote {\bibinfo {title}
      {Anisotropic breakdown of fermi liquid quasiparticle excitations in overdoped
      {La2-xSrxCuO4}},}\ }\href {\doibase 10.1038/ncomms3559} {\bibfield  {journal}
      {\bibinfo  {journal} {Nature Communications}\ }\textbf {\bibinfo {volume}
      {4}},\ \bibinfo {pages} {2559} (\bibinfo {year} {2013})}\BibitemShut
      {NoStop}%
    \bibitem [{\citenamefont {Horio}\ \emph {et~al.}(2020)\citenamefont {Horio},
      \citenamefont {Kramer}, \citenamefont {Wang}, \citenamefont {Zaidan},
      \citenamefont {von Arx}, \citenamefont {Sutter}, \citenamefont {Matt},
      \citenamefont {Sassa}, \citenamefont {Plumb}, \citenamefont {Shi},
      \citenamefont {Hanff}, \citenamefont {Mahatha}, \citenamefont {Bentmann},
      \citenamefont {Reinert}, \citenamefont {Rohlf}, \citenamefont {Diekmann},
      \citenamefont {Buck}, \citenamefont {Kall\"ane}, \citenamefont {Rossnagel},
      \citenamefont {Rienks}, \citenamefont {Granata}, \citenamefont {Fittipaldi},
      \citenamefont {Vecchione}, \citenamefont {Ohgi}, \citenamefont {Kawamata},
      \citenamefont {Adachi}, \citenamefont {Koike}, \citenamefont {Fujimori},
      \citenamefont {Hoesch},\ and\ \citenamefont {Chang}}]{Horio_2020}%
      \BibitemOpen
      \bibfield  {author} {\bibinfo {author} {\bibfnamefont {M.}~\bibnamefont
      {Horio}}, \bibinfo {author} {\bibfnamefont {K.~P.}\ \bibnamefont {Kramer}},
      \bibinfo {author} {\bibfnamefont {Q.}~\bibnamefont {Wang}}, \bibinfo {author}
      {\bibfnamefont {A.}~\bibnamefont {Zaidan}}, \bibinfo {author} {\bibfnamefont
      {K.}~\bibnamefont {von Arx}}, \bibinfo {author} {\bibfnamefont
      {D.}~\bibnamefont {Sutter}}, \bibinfo {author} {\bibfnamefont {C.~E.}\
      \bibnamefont {Matt}}, \bibinfo {author} {\bibfnamefont {Y.}~\bibnamefont
      {Sassa}}, \bibinfo {author} {\bibfnamefont {N.~C.}\ \bibnamefont {Plumb}},
      \bibinfo {author} {\bibfnamefont {M.}~\bibnamefont {Shi}}, \bibinfo {author}
      {\bibfnamefont {A.}~\bibnamefont {Hanff}}, \bibinfo {author} {\bibfnamefont
      {S.~K.}\ \bibnamefont {Mahatha}}, \bibinfo {author} {\bibfnamefont
      {H.}~\bibnamefont {Bentmann}}, \bibinfo {author} {\bibfnamefont
      {F.}~\bibnamefont {Reinert}}, \bibinfo {author} {\bibfnamefont
      {S.}~\bibnamefont {Rohlf}}, \bibinfo {author} {\bibfnamefont {F.~K.}\
      \bibnamefont {Diekmann}}, \bibinfo {author} {\bibfnamefont {J.}~\bibnamefont
      {Buck}}, \bibinfo {author} {\bibfnamefont {M.}~\bibnamefont {Kall\"ane}},
      \bibinfo {author} {\bibfnamefont {K.}~\bibnamefont {Rossnagel}}, \bibinfo
      {author} {\bibfnamefont {E.}~\bibnamefont {Rienks}}, \bibinfo {author}
      {\bibfnamefont {V.}~\bibnamefont {Granata}}, \bibinfo {author} {\bibfnamefont
      {R.}~\bibnamefont {Fittipaldi}}, \bibinfo {author} {\bibfnamefont
      {A.}~\bibnamefont {Vecchione}}, \bibinfo {author} {\bibfnamefont
      {T.}~\bibnamefont {Ohgi}}, \bibinfo {author} {\bibfnamefont {T.}~\bibnamefont
      {Kawamata}}, \bibinfo {author} {\bibfnamefont {T.}~\bibnamefont {Adachi}},
      \bibinfo {author} {\bibfnamefont {Y.}~\bibnamefont {Koike}}, \bibinfo
      {author} {\bibfnamefont {A.}~\bibnamefont {Fujimori}}, \bibinfo {author}
      {\bibfnamefont {M.}~\bibnamefont {Hoesch}}, \ and\ \bibinfo {author}
      {\bibfnamefont {J.}~\bibnamefont {Chang}},\ }\bibfield  {title} {\enquote
      {\bibinfo {title} {Oxide fermi liquid universality revealed by electron
      spectroscopy},}\ }\href {\doibase 10.1103/PhysRevB.102.245153} {\bibfield
      {journal} {\bibinfo  {journal} {Phys. Rev. B}\ }\textbf {\bibinfo {volume}
      {102}},\ \bibinfo {pages} {245153} (\bibinfo {year} {2020})}\BibitemShut
      {NoStop}%
    \bibitem [{\citenamefont {Armitage}\ \emph {et~al.}(2002)\citenamefont
      {Armitage}, \citenamefont {Ronning}, \citenamefont {Lu}, \citenamefont {Kim},
      \citenamefont {Damascelli}, \citenamefont {Shen}, \citenamefont {Feng},
      \citenamefont {Eisaki}, \citenamefont {Shen}, \citenamefont {Mang},\ and\
      \citenamefont {et~al.}}]{Armitage_2002}%
      \BibitemOpen
      \bibfield  {author} {\bibinfo {author} {\bibfnamefont {N.~P.}\ \bibnamefont
      {Armitage}}, \bibinfo {author} {\bibfnamefont {F.}~\bibnamefont {Ronning}},
      \bibinfo {author} {\bibfnamefont {D.~H.}\ \bibnamefont {Lu}}, \bibinfo
      {author} {\bibfnamefont {C.}~\bibnamefont {Kim}}, \bibinfo {author}
      {\bibfnamefont {A.}~\bibnamefont {Damascelli}}, \bibinfo {author}
      {\bibfnamefont {K.~M.}\ \bibnamefont {Shen}}, \bibinfo {author}
      {\bibfnamefont {D.~L.}\ \bibnamefont {Feng}}, \bibinfo {author}
      {\bibfnamefont {H.}~\bibnamefont {Eisaki}}, \bibinfo {author} {\bibfnamefont
      {Z.-X.}\ \bibnamefont {Shen}}, \bibinfo {author} {\bibfnamefont {P.~K.}\
      \bibnamefont {Mang}}, \ and\ \bibinfo {author} {\bibnamefont {et~al.}},\
      }\bibfield  {title} {\enquote {\bibinfo {title} {Doping dependence of an
      n-type cuprate superconductor investigated by angle-resolved photoemission
      spectroscopy},}\ }\href {\doibase 10.1103/PhysRevLett.88.257001} {\bibfield
      {journal} {\bibinfo  {journal} {Physical Review Letters}\ }\textbf {\bibinfo
      {volume} {88}},\ \bibinfo {pages} {257001} (\bibinfo {year}
      {2002})}\BibitemShut {NoStop}%
    \bibitem [{\citenamefont {Motoyama}\ \emph {et~al.}(2007)\citenamefont
      {Motoyama}, \citenamefont {Yu}, \citenamefont {Vishik}, \citenamefont {Vajk},
      \citenamefont {Mang},\ and\ \citenamefont {Greven}}]{Motoyama_2007}%
      \BibitemOpen
      \bibfield  {author} {\bibinfo {author} {\bibfnamefont {E.~M.}\ \bibnamefont
      {Motoyama}}, \bibinfo {author} {\bibfnamefont {G.}~\bibnamefont {Yu}},
      \bibinfo {author} {\bibfnamefont {I.~M.}\ \bibnamefont {Vishik}}, \bibinfo
      {author} {\bibfnamefont {O.~P.}\ \bibnamefont {Vajk}}, \bibinfo {author}
      {\bibfnamefont {P.~K.}\ \bibnamefont {Mang}}, \ and\ \bibinfo {author}
      {\bibfnamefont {M.}~\bibnamefont {Greven}},\ }\bibfield  {title} {\enquote
      {\bibinfo {title} {Spin correlations in the electron-doped
      high-transition-temperature superconductor {Nd 2-x Ce x CuO 4}},}\ }\href
      {\doibase 10.1038/nature05437} {\bibfield  {journal} {\bibinfo  {journal}
      {Nature}\ }\textbf {\bibinfo {volume} {445}},\ \bibinfo {pages} {186–189}
      (\bibinfo {year} {2007})}\BibitemShut {NoStop}%
    \bibitem [{\citenamefont {Armitage}\ \emph {et~al.}(2010)\citenamefont
      {Armitage}, \citenamefont {Fournier},\ and\ \citenamefont
      {Greene}}]{Armitage:2010}%
      \BibitemOpen
      \bibfield  {author} {\bibinfo {author} {\bibfnamefont {N.~P.}\ \bibnamefont
      {Armitage}}, \bibinfo {author} {\bibfnamefont {P.}~\bibnamefont {Fournier}},
      \ and\ \bibinfo {author} {\bibfnamefont {R.~L.}\ \bibnamefont {Greene}},\
      }\bibfield  {title} {\enquote {\bibinfo {title} {Progress and perspectives on
      electron-doped cuprates},}\ }\href {\doibase 10.1103/RevModPhys.82.2421}
      {\bibfield  {journal} {\bibinfo  {journal} {Rev. Mod. Phys.}\ }\textbf
      {\bibinfo {volume} {82}},\ \bibinfo {pages} {2421--2487} (\bibinfo {year}
      {2010})}\BibitemShut {NoStop}%
    \bibitem [{\citenamefont {Berg}\ \emph {et~al.}(2019)\citenamefont {Berg},
      \citenamefont {Lederer}, \citenamefont {Schattner},\ and\ \citenamefont
      {Trebst}}]{Berg_QMC:2019}%
      \BibitemOpen
      \bibfield  {author} {\bibinfo {author} {\bibfnamefont {Erez}\ \bibnamefont
      {Berg}}, \bibinfo {author} {\bibfnamefont {Samuel}\ \bibnamefont {Lederer}},
      \bibinfo {author} {\bibfnamefont {Yoni}\ \bibnamefont {Schattner}}, \ and\
      \bibinfo {author} {\bibfnamefont {Simon}\ \bibnamefont {Trebst}},\ }\bibfield
       {title} {\enquote {\bibinfo {title} {Monte carlo studies of quantum critical
      metals},}\ }\href {\doibase 10.1146/annurev-conmatphys-031218-013339}
      {\bibfield  {journal} {\bibinfo  {journal} {Annual Review of Condensed Matter
      Physics}\ }\textbf {\bibinfo {volume} {10}},\ \bibinfo {pages} {63--84}
      (\bibinfo {year} {2019})},\ \Eprint
      {http://arxiv.org/abs/https://doi.org/10.1146/annurev-conmatphys-031218-013339}
      {https://doi.org/10.1146/annurev-conmatphys-031218-013339} \BibitemShut
      {NoStop}%
    \bibitem [{\citenamefont {Klein}\ \emph {et~al.}(2020)\citenamefont {Klein},
      \citenamefont {Chubukov}, \citenamefont {Schattner},\ and\ \citenamefont
      {Berg}}]{Klein_Chubukov_QMC:2020}%
      \BibitemOpen
      \bibfield  {author} {\bibinfo {author} {\bibfnamefont {Avraham}\ \bibnamefont
      {Klein}}, \bibinfo {author} {\bibfnamefont {Andrey~V.}\ \bibnamefont
      {Chubukov}}, \bibinfo {author} {\bibfnamefont {Yoni}\ \bibnamefont
      {Schattner}}, \ and\ \bibinfo {author} {\bibfnamefont {Erez}\ \bibnamefont
      {Berg}},\ }\bibfield  {title} {\enquote {\bibinfo {title} {Normal state
      properties of quantum critical metals at finite temperature},}\ }\href
      {\doibase 10.1103/PhysRevX.10.031053} {\bibfield  {journal} {\bibinfo
      {journal} {Phys. Rev. X}\ }\textbf {\bibinfo {volume} {10}},\ \bibinfo
      {pages} {031053} (\bibinfo {year} {2020})}\BibitemShut {NoStop}%
    \bibitem [{\citenamefont {Sachdev}\ \emph {et~al.}(1995)\citenamefont
      {Sachdev}, \citenamefont {Chubukov},\ and\ \citenamefont
      {Sokol}}]{Sachdev_1995}%
      \BibitemOpen
      \bibfield  {author} {\bibinfo {author} {\bibfnamefont {Subir}\ \bibnamefont
      {Sachdev}}, \bibinfo {author} {\bibfnamefont {Andrey~V.}\ \bibnamefont
      {Chubukov}}, \ and\ \bibinfo {author} {\bibfnamefont {Alexander}\
      \bibnamefont {Sokol}},\ }\bibfield  {title} {\enquote {\bibinfo {title}
      {Crossover and scaling in a nearly antiferromagnetic fermi liquid in two
      dimensions},}\ }\href {\doibase 10.1103/PhysRevB.51.14874} {\bibfield
      {journal} {\bibinfo  {journal} {Phys. Rev. B}\ }\textbf {\bibinfo {volume}
      {51}},\ \bibinfo {pages} {14874--14891} (\bibinfo {year} {1995})}\BibitemShut
      {NoStop}%
    \bibitem [{\citenamefont {Hlubina}\ and\ \citenamefont
      {Rice}(1995)}]{Hlubina:1995}%
      \BibitemOpen
      \bibfield  {author} {\bibinfo {author} {\bibfnamefont {R.}~\bibnamefont
      {Hlubina}}\ and\ \bibinfo {author} {\bibfnamefont {T.~M.}\ \bibnamefont
      {Rice}},\ }\bibfield  {title} {\enquote {\bibinfo {title} {Erratum:
      Resistivity as a function of temperature for models with hot spots on the
      fermi surface},}\ }\href {\doibase 10.1103/PhysRevB.52.13043.3} {\bibfield
      {journal} {\bibinfo  {journal} {Phys. Rev. B}\ }\textbf {\bibinfo {volume}
      {52}},\ \bibinfo {pages} {13043--13043} (\bibinfo {year} {1995})}\BibitemShut
      {NoStop}%
    \bibitem [{\citenamefont {L\"ohneysen}\ \emph
      {et~al.}(2007{\natexlab{b}})\citenamefont {L\"ohneysen}, \citenamefont
      {Rosch}, \citenamefont {Vojta},\ and\ \citenamefont
      {W\"olfle}}]{Lohneysen_2007}%
      \BibitemOpen
      \bibfield  {author} {\bibinfo {author} {\bibfnamefont {Hilbert~v.}\
      \bibnamefont {L\"ohneysen}}, \bibinfo {author} {\bibfnamefont {Achim}\
      \bibnamefont {Rosch}}, \bibinfo {author} {\bibfnamefont {Matthias}\
      \bibnamefont {Vojta}}, \ and\ \bibinfo {author} {\bibfnamefont {Peter}\
      \bibnamefont {W\"olfle}},\ }\bibfield  {title} {\enquote {\bibinfo {title}
      {Fermi-liquid instabilities at magnetic quantum phase transitions},}\ }\href
      {\doibase 10.1103/RevModPhys.79.1015} {\bibfield  {journal} {\bibinfo
      {journal} {Rev. Mod. Phys.}\ }\textbf {\bibinfo {volume} {79}},\ \bibinfo
      {pages} {1015--1075} (\bibinfo {year} {2007}{\natexlab{b}})}\BibitemShut
      {NoStop}%
    \bibitem [{\citenamefont {Basov}\ \emph {et~al.}(2011)\citenamefont {Basov},
      \citenamefont {Averitt}, \citenamefont {van~der Marel}, \citenamefont
      {Dressel},\ and\ \citenamefont
      {Haule}}]{Basov_Averitt_van_der_Marel_Dressel_Haule_2011}%
      \BibitemOpen
      \bibfield  {author} {\bibinfo {author} {\bibfnamefont {D.~N.}\ \bibnamefont
      {Basov}}, \bibinfo {author} {\bibfnamefont {Richard~D.}\ \bibnamefont
      {Averitt}}, \bibinfo {author} {\bibfnamefont {Dirk}\ \bibnamefont {van~der
      Marel}}, \bibinfo {author} {\bibfnamefont {Martin}\ \bibnamefont {Dressel}},
      \ and\ \bibinfo {author} {\bibfnamefont {Kristjan}\ \bibnamefont {Haule}},\
      }\bibfield  {title} {\enquote {\bibinfo {title} {Electrodynamics of
      correlated electron materials},}\ }\href {\doibase 10.1103/RevModPhys.83.471}
      {\bibfield  {journal} {\bibinfo  {journal} {Reviews of Modern Physics}\
      }\textbf {\bibinfo {volume} {83}},\ \bibinfo {pages} {471–541} (\bibinfo
      {year} {2011})}\BibitemShut {NoStop}%
    \bibitem [{\citenamefont {Matsui}\ \emph {et~al.}(2005)\citenamefont {Matsui},
      \citenamefont {Terashima}, \citenamefont {Sato}, \citenamefont {Takahashi},
      \citenamefont {Wang}, \citenamefont {Yang}, \citenamefont {Ding},
      \citenamefont {Uefuji},\ and\ \citenamefont {Yamada}}]{Matsui_2005}%
      \BibitemOpen
      \bibfield  {author} {\bibinfo {author} {\bibfnamefont {H.}~\bibnamefont
      {Matsui}}, \bibinfo {author} {\bibfnamefont {K.}~\bibnamefont {Terashima}},
      \bibinfo {author} {\bibfnamefont {T.}~\bibnamefont {Sato}}, \bibinfo {author}
      {\bibfnamefont {T.}~\bibnamefont {Takahashi}}, \bibinfo {author}
      {\bibfnamefont {S.-C.}\ \bibnamefont {Wang}}, \bibinfo {author}
      {\bibfnamefont {H.-B.}\ \bibnamefont {Yang}}, \bibinfo {author}
      {\bibfnamefont {H.}~\bibnamefont {Ding}}, \bibinfo {author} {\bibfnamefont
      {T.}~\bibnamefont {Uefuji}}, \ and\ \bibinfo {author} {\bibfnamefont
      {K.}~\bibnamefont {Yamada}},\ }\bibfield  {title} {\enquote {\bibinfo {title}
      {Angle-resolved photoemission spectroscopy of the antiferromagnetic
      superconductor nd1.87ce0.13cuo4: Anisotropic spin-correlation gap, pseudogap,
      and the induced quasiparticle mass enhancement},}\ }\href {\doibase
      10.1103/PhysRevLett.94.047005} {\bibfield  {journal} {\bibinfo  {journal}
      {Physical Review Letters}\ }\textbf {\bibinfo {volume} {94}},\ \bibinfo
      {pages} {047005} (\bibinfo {year} {2005})}\BibitemShut {NoStop}%
    \bibitem [{\citenamefont {Matsui}\ \emph {et~al.}(2006)\citenamefont {Matsui},
      \citenamefont {Terashima}, \citenamefont {Sato}, \citenamefont {Takahashi},
      \citenamefont {Wang}, \citenamefont {Yang}, \citenamefont {Ding},
      \citenamefont {Uefuji},\ and\ \citenamefont {Yamada}}]{Matsui_2006}%
      \BibitemOpen
      \bibfield  {author} {\bibinfo {author} {\bibfnamefont {Hiroaki}\ \bibnamefont
      {Matsui}}, \bibinfo {author} {\bibfnamefont {Kensei}\ \bibnamefont
      {Terashima}}, \bibinfo {author} {\bibfnamefont {Takafumi}\ \bibnamefont
      {Sato}}, \bibinfo {author} {\bibfnamefont {Takashi}\ \bibnamefont
      {Takahashi}}, \bibinfo {author} {\bibfnamefont {Shancai}\ \bibnamefont
      {Wang}}, \bibinfo {author} {\bibfnamefont {Hongbo}\ \bibnamefont {Yang}},
      \bibinfo {author} {\bibfnamefont {Hong}\ \bibnamefont {Ding}}, \bibinfo
      {author} {\bibfnamefont {Tetsushi}\ \bibnamefont {Uefuji}}, \ and\ \bibinfo
      {author} {\bibfnamefont {Kazuyoshi}\ \bibnamefont {Yamada}},\ }\bibfield
      {title} {\enquote {\bibinfo {title} {Arpes study of quasiparticle state in
      electron-doped cuprate nd2cecuo4},}\ }\href {\doibase
      10.1016/j.jpcs.2005.10.036} {\bibfield  {journal} {\bibinfo  {journal}
      {Journal of Physics and Chemistry of Solids}\ }\textbf {\bibinfo {volume}
      {67}},\ \bibinfo {pages} {249–253} (\bibinfo {year} {2006})}\BibitemShut
      {NoStop}%
    \bibitem [{\citenamefont {He}\ \emph {et~al.}(2019)\citenamefont {He},
      \citenamefont {Rotundu}, \citenamefont {Scheurer}, \citenamefont {He},
      \citenamefont {Hashimoto}, \citenamefont {Xu}, \citenamefont {Wang},
      \citenamefont {Huang}, \citenamefont {Jia}, \citenamefont {Chen},\ and\
      \citenamefont {et~al.}}]{He_2019}%
      \BibitemOpen
      \bibfield  {author} {\bibinfo {author} {\bibfnamefont {Junfeng}\ \bibnamefont
      {He}}, \bibinfo {author} {\bibfnamefont {Costel~R.}\ \bibnamefont {Rotundu}},
      \bibinfo {author} {\bibfnamefont {Mathias~S.}\ \bibnamefont {Scheurer}},
      \bibinfo {author} {\bibfnamefont {Yu}~\bibnamefont {He}}, \bibinfo {author}
      {\bibfnamefont {Makoto}\ \bibnamefont {Hashimoto}}, \bibinfo {author}
      {\bibfnamefont {Ke-Jun}\ \bibnamefont {Xu}}, \bibinfo {author} {\bibfnamefont
      {Yao}\ \bibnamefont {Wang}}, \bibinfo {author} {\bibfnamefont {Edwin~W.}\
      \bibnamefont {Huang}}, \bibinfo {author} {\bibfnamefont {Tao}\ \bibnamefont
      {Jia}}, \bibinfo {author} {\bibfnamefont {Sudi}\ \bibnamefont {Chen}}, \ and\
      \bibinfo {author} {\bibnamefont {et~al.}},\ }\bibfield  {title} {\enquote
      {\bibinfo {title} {Fermi surface reconstruction in electron-doped cuprates
      without antiferromagnetic long-range order},}\ }\href {\doibase
      10.1073/pnas.1816121116} {\bibfield  {journal} {\bibinfo  {journal}
      {Proceedings of the National Academy of Sciences}\ }\textbf {\bibinfo
      {volume} {116}},\ \bibinfo {pages} {3449–3453} (\bibinfo {year}
      {2019})}\BibitemShut {NoStop}%
    \bibitem [{\citenamefont {Kyung}\ \emph {et~al.}(2004)\citenamefont {Kyung},
      \citenamefont {Hankevych}, \citenamefont {Daré},\ and\ \citenamefont
      {Tremblay}}]{Kyung_2004}%
      \BibitemOpen
      \bibfield  {author} {\bibinfo {author} {\bibfnamefont {B.}~\bibnamefont
      {Kyung}}, \bibinfo {author} {\bibfnamefont {V.}~\bibnamefont {Hankevych}},
      \bibinfo {author} {\bibfnamefont {A.-M.}\ \bibnamefont {Daré}}, \ and\
      \bibinfo {author} {\bibfnamefont {A.-M.~S.}\ \bibnamefont {Tremblay}},\
      }\bibfield  {title} {\enquote {\bibinfo {title} {Pseudogap and spin
      fluctuations in the normal state of the electron-doped cuprates},}\ }\href
      {\doibase 10.1103/PhysRevLett.93.147004} {\bibfield  {journal} {\bibinfo
      {journal} {Physical Review Letters}\ }\textbf {\bibinfo {volume} {93}},\
      \bibinfo {pages} {147004} (\bibinfo {year} {2004})}\BibitemShut {NoStop}%
    \bibitem [{\citenamefont {Schäfer}\ \emph {et~al.}(2021)\citenamefont
      {Schäfer}, \citenamefont {Wentzell}, \citenamefont {Šimkovic},
      \citenamefont {He}, \citenamefont {Hille}, \citenamefont {Klett},
      \citenamefont {Eckhardt}, \citenamefont {Arzhang}, \citenamefont {Harkov},
      \citenamefont {Le~Régent},\ and\ \citenamefont {et~al.}}]{Schafer_2021}%
      \BibitemOpen
      \bibfield  {author} {\bibinfo {author} {\bibfnamefont {Thomas}\ \bibnamefont
      {Schäfer}}, \bibinfo {author} {\bibfnamefont {Nils}\ \bibnamefont
      {Wentzell}}, \bibinfo {author} {\bibfnamefont {Fedor}\ \bibnamefont
      {Šimkovic}}, \bibinfo {author} {\bibfnamefont {Yuan-Yao}\ \bibnamefont
      {He}}, \bibinfo {author} {\bibfnamefont {Cornelia}\ \bibnamefont {Hille}},
      \bibinfo {author} {\bibfnamefont {Marcel}\ \bibnamefont {Klett}}, \bibinfo
      {author} {\bibfnamefont {Christian~J.}\ \bibnamefont {Eckhardt}}, \bibinfo
      {author} {\bibfnamefont {Behnam}\ \bibnamefont {Arzhang}}, \bibinfo {author}
      {\bibfnamefont {Viktor}\ \bibnamefont {Harkov}}, \bibinfo {author}
      {\bibfnamefont {François-Marie}\ \bibnamefont {Le~Régent}}, \ and\ \bibinfo
      {author} {\bibnamefont {et~al.}},\ }\bibfield  {title} {\enquote {\bibinfo
      {title} {Tracking the footprints of spin fluctuations: A multimethod,
      multimessenger study of the two-dimensional hubbard model},}\ }\href
      {\doibase 10.1103/PhysRevX.11.011058} {\bibfield  {journal} {\bibinfo
      {journal} {Physical Review X}\ }\textbf {\bibinfo {volume} {11}},\ \bibinfo
      {pages} {011058} (\bibinfo {year} {2021})}\BibitemShut {NoStop}%
    \bibitem [{\citenamefont {Boschini}\ \emph {et~al.}(2020)\citenamefont
      {Boschini}, \citenamefont {Zonno}, \citenamefont {Razzoli}, \citenamefont
      {Day}, \citenamefont {Michiardi}, \citenamefont {Zwartsenberg}, \citenamefont
      {Nigge}, \citenamefont {Schneider}, \citenamefont {da~Silva~Neto},
      \citenamefont {Erb},\ and\ \citenamefont {et~al.}}]{Boschini_2020}%
      \BibitemOpen
      \bibfield  {author} {\bibinfo {author} {\bibfnamefont {Fabio}\ \bibnamefont
      {Boschini}}, \bibinfo {author} {\bibfnamefont {Marta}\ \bibnamefont {Zonno}},
      \bibinfo {author} {\bibfnamefont {Elia}\ \bibnamefont {Razzoli}}, \bibinfo
      {author} {\bibfnamefont {Ryan~P.}\ \bibnamefont {Day}}, \bibinfo {author}
      {\bibfnamefont {Matteo}\ \bibnamefont {Michiardi}}, \bibinfo {author}
      {\bibfnamefont {Berend}\ \bibnamefont {Zwartsenberg}}, \bibinfo {author}
      {\bibfnamefont {Pascal}\ \bibnamefont {Nigge}}, \bibinfo {author}
      {\bibfnamefont {Michael}\ \bibnamefont {Schneider}}, \bibinfo {author}
      {\bibfnamefont {Eduardo~H.}\ \bibnamefont {da~Silva~Neto}}, \bibinfo {author}
      {\bibfnamefont {Andreas}\ \bibnamefont {Erb}}, \ and\ \bibinfo {author}
      {\bibnamefont {et~al.}},\ }\bibfield  {title} {\enquote {\bibinfo {title}
      {Emergence of pseudogap from short-range spin-correlations in electron-doped
      cuprates},}\ }\href {\doibase 10.1038/s41535-020-0208-6} {\bibfield
      {journal} {\bibinfo  {journal} {npj Quantum Materials}\ }\textbf {\bibinfo
      {volume} {5}},\ \bibinfo {pages} {6} (\bibinfo {year} {2020})}\BibitemShut
      {NoStop}%
    \bibitem [{\citenamefont {Horio}\ \emph {et~al.}(2016)\citenamefont {Horio},
      \citenamefont {Adachi}, \citenamefont {Mori}, \citenamefont {Takahashi},
      \citenamefont {Yoshida}, \citenamefont {Suzuki}, \citenamefont {Ambolode},
      \citenamefont {Okazaki}, \citenamefont {Ono}, \citenamefont {Kumigashira},
      \citenamefont {Anzai}, \citenamefont {Arita}, \citenamefont {Namatame},
      \citenamefont {Taniguchi}, \citenamefont {Ootsuki}, \citenamefont {Sawada},
      \citenamefont {Takahashi}, \citenamefont {Mizokawa}, \citenamefont {Koike},\
      and\ \citenamefont {Fujimori}}]{Horio_2016}%
      \BibitemOpen
      \bibfield  {author} {\bibinfo {author} {\bibfnamefont {M.}~\bibnamefont
      {Horio}}, \bibinfo {author} {\bibfnamefont {T.}~\bibnamefont {Adachi}},
      \bibinfo {author} {\bibfnamefont {Y.}~\bibnamefont {Mori}}, \bibinfo {author}
      {\bibfnamefont {A.}~\bibnamefont {Takahashi}}, \bibinfo {author}
      {\bibfnamefont {T.}~\bibnamefont {Yoshida}}, \bibinfo {author} {\bibfnamefont
      {H.}~\bibnamefont {Suzuki}}, \bibinfo {author} {\bibfnamefont {L.~C.~C.}\
      \bibnamefont {Ambolode}}, \bibinfo {author} {\bibfnamefont {K.}~\bibnamefont
      {Okazaki}}, \bibinfo {author} {\bibfnamefont {K.}~\bibnamefont {Ono}},
      \bibinfo {author} {\bibfnamefont {H.}~\bibnamefont {Kumigashira}}, \bibinfo
      {author} {\bibfnamefont {H.}~\bibnamefont {Anzai}}, \bibinfo {author}
      {\bibfnamefont {M.}~\bibnamefont {Arita}}, \bibinfo {author} {\bibfnamefont
      {H.}~\bibnamefont {Namatame}}, \bibinfo {author} {\bibfnamefont
      {M.}~\bibnamefont {Taniguchi}}, \bibinfo {author} {\bibfnamefont
      {D.}~\bibnamefont {Ootsuki}}, \bibinfo {author} {\bibfnamefont
      {K.}~\bibnamefont {Sawada}}, \bibinfo {author} {\bibfnamefont
      {M.}~\bibnamefont {Takahashi}}, \bibinfo {author} {\bibfnamefont
      {T.}~\bibnamefont {Mizokawa}}, \bibinfo {author} {\bibfnamefont
      {Y.}~\bibnamefont {Koike}}, \ and\ \bibinfo {author} {\bibfnamefont
      {A.}~\bibnamefont {Fujimori}},\ }\bibfield  {title} {\enquote {\bibinfo
      {title} {Suppression of the antiferromagnetic pseudogap in the electron-doped
      high-temperature superconductor by protect annealing},}\ }\href {\doibase
      10.1038/ncomms10567} {\bibfield  {journal} {\bibinfo  {journal} {Nature
      Communications}\ }\textbf {\bibinfo {volume} {7}},\ \bibinfo {pages} {10567}
      (\bibinfo {year} {2016})}\BibitemShut {NoStop}%
    \bibitem [{\citenamefont {Vilk}\ and\ \citenamefont
      {Tremblay}(1997)}]{Vilk_1997}%
      \BibitemOpen
      \bibfield  {author} {\bibinfo {author} {\bibfnamefont {Y.~M.}\ \bibnamefont
      {Vilk}}\ and\ \bibinfo {author} {\bibfnamefont {A.-M.~S.}\ \bibnamefont
      {Tremblay}},\ }\bibfield  {title} {\enquote {\bibinfo {title}
      {Non-perturbative many-body approach to the hubbard model and single-particle
      pseudogap},}\ }\href {\doibase 10.1051/jp1:1997135} {\bibfield  {journal}
      {\bibinfo  {journal} {Journal de Physique I}\ }\textbf {\bibinfo {volume}
      {7}},\ \bibinfo {pages} {1309–1368} (\bibinfo {year} {1997})},\ \bibinfo
      {note} {arXiv: cond-mat/9702188}\BibitemShut {NoStop}%
    \bibitem [{\citenamefont {Tremblay}(2011)}]{TremblayMancini:2011}%
      \BibitemOpen
      \bibfield  {author} {\bibinfo {author} {\bibfnamefont {A.~M.~S.}\
      \bibnamefont {Tremblay}},\ }\bibfield  {title} {\enquote {\bibinfo {title}
      {Two-particle-self-consistent approach for the hubbard model},}\ }in\
      \href@noop {} {\emph {\bibinfo {booktitle} {Strongly Correlated Systems:
      Theoretical Methods}}},\ \bibinfo {editor} {edited by\ \bibinfo {editor}
      {\bibfnamefont {F.}~\bibnamefont {Mancini}}\ and\ \bibinfo {editor}
      {\bibfnamefont {A.}~\bibnamefont {Avella}}}\ (\bibinfo  {publisher} {Springer
      series},\ \bibinfo {year} {2011})\ Chap.~\bibinfo {chapter} {13}, pp.\
      \bibinfo {pages} {409--455}\BibitemShut {NoStop}%
    \bibitem [{\citenamefont {Kyung}\ \emph {et~al.}(2001)\citenamefont {Kyung},
      \citenamefont {Allen},\ and\ \citenamefont {Tremblay}}]{Kyung_2001}%
      \BibitemOpen
      \bibfield  {author} {\bibinfo {author} {\bibfnamefont {B.}~\bibnamefont
      {Kyung}}, \bibinfo {author} {\bibfnamefont {S.}~\bibnamefont {Allen}}, \ and\
      \bibinfo {author} {\bibfnamefont {A.-M.~S.}\ \bibnamefont {Tremblay}},\
      }\bibfield  {title} {\enquote {\bibinfo {title} {Pairing fluctuations and
      pseudogaps in the attractive hubbard model},}\ }\href {\doibase
      10.1103/PhysRevB.64.075116} {\bibfield  {journal} {\bibinfo  {journal}
      {Physical Review B}\ }\textbf {\bibinfo {volume} {64}},\ \bibinfo {pages}
      {075116} (\bibinfo {year} {2001})}\BibitemShut {NoStop}%
    \bibitem [{\citenamefont {Bergeron}\ \emph {et~al.}(2012)\citenamefont
      {Bergeron}, \citenamefont {Chowdhury}, \citenamefont {Punk}, \citenamefont
      {Sachdev},\ and\ \citenamefont {Tremblay}}]{Bergeron_2012}%
      \BibitemOpen
      \bibfield  {author} {\bibinfo {author} {\bibfnamefont {Dominic}\ \bibnamefont
      {Bergeron}}, \bibinfo {author} {\bibfnamefont {Debanjan}\ \bibnamefont
      {Chowdhury}}, \bibinfo {author} {\bibfnamefont {Matthias}\ \bibnamefont
      {Punk}}, \bibinfo {author} {\bibfnamefont {Subir}\ \bibnamefont {Sachdev}}, \
      and\ \bibinfo {author} {\bibfnamefont {A.-M.~S.}\ \bibnamefont {Tremblay}},\
      }\bibfield  {title} {\enquote {\bibinfo {title} {Breakdown of fermi liquid
      behavior at the (pi,pi) = 2kf spin-density wave quantum-critical point: The
      case of electron-doped cuprates},}\ }\href {\doibase
      10.1103/PhysRevB.86.155123} {\bibfield  {journal} {\bibinfo  {journal}
      {Physical Review B}\ }\textbf {\bibinfo {volume} {86}},\ \bibinfo {pages}
      {155123} (\bibinfo {year} {2012})}\BibitemShut {NoStop}%
    \bibitem [{Note1()}]{Note1}%
      \BibitemOpen
      \bibinfo {note} {See the Supplemental Material at [URL] for the following:
      Fitting procedure for the Matsubara Self-Energy~\cite {Hodges_1971}; Finding
      the exponent for the temperature dependence of the scattering rate using a
      scaling function; The relation between strength of interaction as measured by
      $U$ and $Z$ and cutoff frequency; Discussion of spectroscopic generalizations
      of the Kadowaki-Woods ratio~\cite {Jacko_2009}.}\BibitemShut {Stop}%
    \bibitem [{\citenamefont {Vilk}\ and\ \citenamefont
      {Tremblay}(1996)}]{Vilk:1996}%
      \BibitemOpen
      \bibfield  {author} {\bibinfo {author} {\bibfnamefont {Y.~M.}\ \bibnamefont
      {Vilk}}\ and\ \bibinfo {author} {\bibfnamefont {A.-M.~S.}\ \bibnamefont
      {Tremblay}},\ }\bibfield  {title} {\enquote {\bibinfo {title} {Destruction of
      fermi-liquid quasiparticles in two dimensions by critical fluctuations},}\
      }\href {http://stacks.iop.org/0295-5075/33/i=2/a=159} {\bibfield  {journal}
      {\bibinfo  {journal} {EPL (Europhysics Letters)}\ }\textbf {\bibinfo {volume}
      {33}},\ \bibinfo {pages} {159} (\bibinfo {year} {1996})}\BibitemShut
      {NoStop}%
    \bibitem [{\citenamefont {Sénéchal}\ and\ \citenamefont
      {Tremblay}(2004)}]{Senechal_2004}%
      \BibitemOpen
      \bibfield  {author} {\bibinfo {author} {\bibfnamefont {David}\ \bibnamefont
      {Sénéchal}}\ and\ \bibinfo {author} {\bibfnamefont {A.-M.~S.}\ \bibnamefont
      {Tremblay}},\ }\bibfield  {title} {\enquote {\bibinfo {title} {Hot spots and
      pseudogaps for hole- and electron-doped high-temperature superconductors},}\
      }\href {\doibase 10.1103/PhysRevLett.92.126401} {\bibfield  {journal}
      {\bibinfo  {journal} {Physical Review Letters}\ }\textbf {\bibinfo {volume}
      {92}},\ \bibinfo {pages} {126401} (\bibinfo {year} {2004})}\BibitemShut
      {NoStop}%
    \bibitem [{\citenamefont {Hankevych}\ \emph {et~al.}(2006)\citenamefont
      {Hankevych}, \citenamefont {Kyung}, \citenamefont {Daré}, \citenamefont
      {Sénéchal},\ and\ \citenamefont {Tremblay}}]{Hankevych_2006}%
      \BibitemOpen
      \bibfield  {author} {\bibinfo {author} {\bibfnamefont {V.}~\bibnamefont
      {Hankevych}}, \bibinfo {author} {\bibfnamefont {B.}~\bibnamefont {Kyung}},
      \bibinfo {author} {\bibfnamefont {A.-M.}\ \bibnamefont {Daré}}, \bibinfo
      {author} {\bibfnamefont {D.}~\bibnamefont {Sénéchal}}, \ and\ \bibinfo
      {author} {\bibfnamefont {A.-M.S.}\ \bibnamefont {Tremblay}},\ }\bibfield
      {title} {\enquote {\bibinfo {title} {Strong- and weak-coupling mechanisms for
      pseudogap in electron-doped cuprates},}\ }\href {\doibase
      10.1016/j.jpcs.2005.10.121} {\bibfield  {journal} {\bibinfo  {journal}
      {Journal of Physics and Chemistry of Solids}\ }\textbf {\bibinfo {volume}
      {67}},\ \bibinfo {pages} {189–192} (\bibinfo {year} {2006})}\BibitemShut
      {NoStop}%
    \bibitem [{\citenamefont {Weber}\ \emph {et~al.}(2010)\citenamefont {Weber},
      \citenamefont {Haule},\ and\ \citenamefont {Kotliar}}]{Weber_2010}%
      \BibitemOpen
      \bibfield  {author} {\bibinfo {author} {\bibfnamefont {Cédric}\ \bibnamefont
      {Weber}}, \bibinfo {author} {\bibfnamefont {Kristjan}\ \bibnamefont {Haule}},
      \ and\ \bibinfo {author} {\bibfnamefont {Gabriel}\ \bibnamefont {Kotliar}},\
      }\bibfield  {title} {\enquote {\bibinfo {title} {Strength of correlations in
      electron- and hole-doped cuprates},}\ }\href {\doibase 10.1038/nphys1706}
      {\bibfield  {journal} {\bibinfo  {journal} {Nature Physics}\ }\textbf
      {\bibinfo {volume} {6}},\ \bibinfo {pages} {574–578} (\bibinfo {year}
      {2010})}\BibitemShut {NoStop}%
    \bibitem [{\citenamefont {Hodges}\ \emph {et~al.}(1971)\citenamefont {Hodges},
      \citenamefont {Smith},\ and\ \citenamefont {Wilkins}}]{Hodges_1971}%
      \BibitemOpen
      \bibfield  {author} {\bibinfo {author} {\bibfnamefont {Christopher}\
      \bibnamefont {Hodges}}, \bibinfo {author} {\bibfnamefont {Henrik}\
      \bibnamefont {Smith}}, \ and\ \bibinfo {author} {\bibfnamefont {J.~W.}\
      \bibnamefont {Wilkins}},\ }\bibfield  {title} {\enquote {\bibinfo {title}
      {Effect of fermi surface geometry on electron-electron scattering},}\ }\href
      {\doibase 10.1103/PhysRevB.4.302} {\bibfield  {journal} {\bibinfo  {journal}
      {Physical Review B}\ }\textbf {\bibinfo {volume} {4}},\ \bibinfo {pages}
      {302–311} (\bibinfo {year} {1971})}\BibitemShut {NoStop}%
    \bibitem [{\citenamefont {Jacko}\ \emph {et~al.}(2009)\citenamefont {Jacko},
      \citenamefont {Fjærestad},\ and\ \citenamefont {Powell}}]{Jacko_2009}%
      \BibitemOpen
      \bibfield  {author} {\bibinfo {author} {\bibfnamefont {A.~C.}\ \bibnamefont
      {Jacko}}, \bibinfo {author} {\bibfnamefont {J.~O.}\ \bibnamefont
      {Fjærestad}}, \ and\ \bibinfo {author} {\bibfnamefont {B.~J.}\ \bibnamefont
      {Powell}},\ }\bibfield  {title} {\enquote {\bibinfo {title} {A unified
      explanation of the kadowaki–woods ratio in strongly correlated metals},}\
      }\href {\doibase 10.1038/nphys1249} {\bibfield  {journal} {\bibinfo
      {journal} {Nature Physics}\ }\textbf {\bibinfo {volume} {5}},\ \bibinfo
      {pages} {422–425} (\bibinfo {year} {2009})}\BibitemShut {NoStop}%
    \end{thebibliography}

%


\setcounter{equation}{0}
\setcounter{figure}{0}
\setcounter{table}{0}
\setcounter{page}{1}
\makeatletter

\renewcommand{\theequation}{S\arabic{equation}}
\renewcommand{\thefigure}{S\arabic{figure}}
\renewcommand{\thetable}{S\arabic{table}}
\renewcommand{\bibnumfmt}[1]{[S#1]}
\renewcommand{\citenumfont}[1]{S#1}

\clearpage

\begin{appendix}

    \begin{center}
        \textbf{\large SUPPLEMENTAL MATERIAL}
        \linebreak
    
        \textbf{\large Resilient Fermi liquid and strength of correlations near a quantum critical point}
        \linebreak
    
        C. Gauvin-Ndiaye, M. Setrakian, and A.-M. S. Tremblay
    \end{center}

    In this Supplemental Material, we discuss in turn
    \begin{itemize}
        \item The fitting procedure for the Matsubara self-energy, 
        \item How to find the exponent for the temperature dependence of the scattering rate as a function of temperature at $\omega=0$ using a scaling function, 
        \item The relation between strength of interaction as measured by $U$ and $Z$ and cutoff frequency, 
        \item Spectroscopic generalizations of the Kadowaki-Woods ratio.
    \end{itemize}

    \section{Fitting the Matsubara Self-Energy} 

    Because of the simple relation between $\Sigma''$ \eref{eq:sigma_reel} and the corresponding Matsubara expression \eref{eq:sigma_mats} at low frequency, one can extract the real-frequency self-energy parameters  $a_0(T)$, $a_1$  and $a_2$ from a polynomial fit of the imaginary-frequency self-energy data that we compute using the TPSC approach. However, this is not straightforward due to the discrete nature of the Matsubara frequencies and their temperature dependence. To avoid these issues, we consider the expected temperature dependence of the imaginary part of the retarded self-energy for a Fermi liquid, which is (neglecting logarithmic corrections which should appear in a 2d system with a cylindrical Fermi surface \cite{Hodges_1971}). 
    \begin{equation}
    \Sigma''(\omega,T) = -a (\pi T)^2 -a \omega^2.
    \label{eq:sigma_T}
    \end{equation}
    In the Matsubara frequency formulation, this yields, for the first imaginary frequency $\omega_0$,
    \begin{equation}
    \mathrm{Im}\Sigma(\omega_0,T) = a_1 \pi T.
    \label{eq:a1T}
    \end{equation}
    
\begin{figure}
  \centering
  \includegraphics[width=\columnwidth]{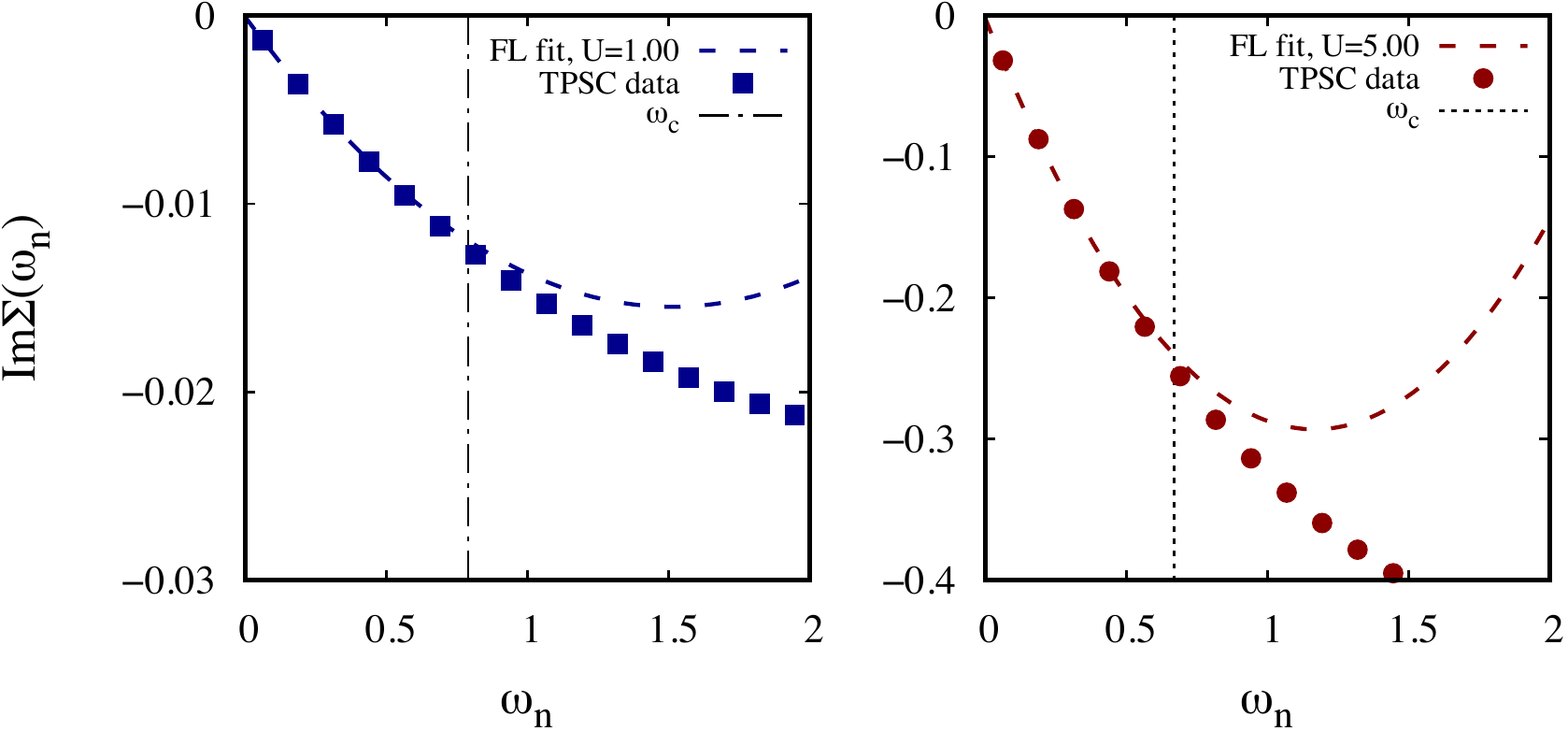} 
  \caption{Fermi-liquid fits of the imaginary part of the self energy in Matsubara frequencies for $x=0.20$ at the antinode, with $U=1$ (left) and $U=5$ (right).}
  \label{fig:Fit_and_wc}
  \end{figure}

    Hence, our first step in the computation of the fit parameters is to extract the parameter $a_1$ through a fit of $\mathrm{Im}\Sigma(\omega_0,T)$ as a function of temperature. Then, we calculate the remaining fit parameters $a_0(T)$ and $a_2$ with a fit of
    \begin{equation}\label{eq:SigmaMatsubara}
    \mathrm{Im}\Sigma(\omega_n,T) - a_1\omega_n = a_0(T)+ a_2 \omega_n^2.
    \end{equation}
    We perform this fit over the first two Matsubara frequencies for multiple temperatures. We obtain the final value of the parameter $a_2$ by taking the average of the computed values for each temperature. We emphasize that this fitting method, and the resulting values of $Z$ and $a_2$ shown in the main text, assume that the self-energy takes a Fermi-liquid form. Examples of the resulting fits are shown in \fref{fig:Fit_and_wc}.

\begin{figure}
    \centering
    \includegraphics[width=\columnwidth]{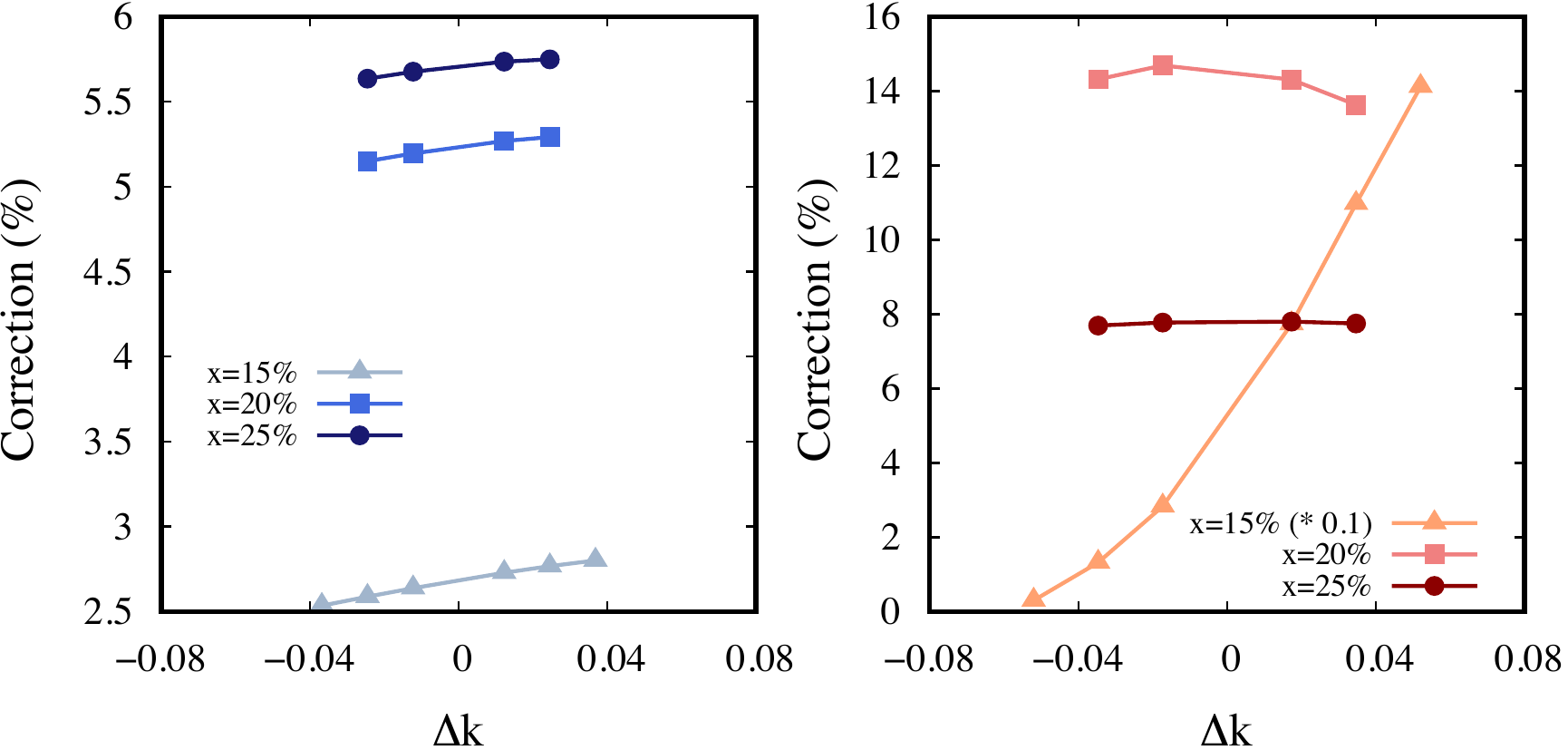} 
    \caption{Correction to the effective mass as defined by \eref{eq:mass_supp} as a function of doping and of the value of the finite difference $\Delta k$. The left panel shows the correction at the antinode, while the right panel shows the correction at the node. The correction at the node for $x=0.15$ is divided by $10$ in the plot to illustrate it on the same axis as the other dopings.}
    \label{fig:corrections}
    \end{figure}

    To obtain the correction to the effective mass
    \begin{equation}
    \frac{\partial \Sigma'(\mathbf{k},\omega=0)}{\partial \xi_{\mathbf{k}}} = \frac{\hat{e}_{\mathbf{k}}\cdot \nabla_{\mathbf{k}}\Sigma'(\mathbf{k},\omega=0)}{\hat{e}_{\mathbf{k}}\cdot \nabla_{\mathbf{k}}\xi_{\mathbf{k}}},
    \label{eq:mass_supp}
    \end{equation}
    we do a second-degree polynomial fit over the first three Matsubara frequency of the real part of the Matsubara self-energy to extract an approximate value of $\Sigma'(\mathbf{k},\omega=0)$. We do this for multiple values of the wavevector $\mathbf{k}$ in directions perpendicular to the node and the antinode. Then, we compute the self-energy gradients with finite differences 
    \begin{equation}
    \nabla_{\mathbf{k}} \Sigma'(\mathbf{k}) \simeq (\Sigma'(\mathbf{k}\pm\Delta k) - \Sigma'(\mathbf{k}) )/\Delta k.
    \label{eq:finite_diff}
    \end{equation}
    The values we obtain from \eref{eq:mass_supp} are shown in \fref{fig:corrections}, where we see that there is a large discrepancy between the left and right derivatives at the node for $x=0.15$. 

    \section{Procedure to extract the exponent for the temperature dependence of the $\omega=0$ imaginary part of the self-energy}

    Assuming, as in Ref.~\cite{Schafer_2021}, that the scaling form 
    \begin{equation}
        \Sigma''(\omega) = a(\pi T)^{\nu}\phi\left(\frac{\omega}{\pi T}\right)
        \label{eq:scalingSupp}    
    \end{equation}
    in \eref{eq:scaling} is valid, the temperature dependence $T^{\nu}$ of the imaginary part of the self-energy can be deduced from a fit of the Matsubara self-energy. To do so, we use the spectral representation of the self-energy:
\begin{equation}
\Sigma(i\omega_n) = \int \frac{d\omega}{\pi} \frac{\Sigma''(\omega)}{\omega - i\omega_n}.
\label{eq:spectral_self}
\end{equation}
Then, we perform a linear fit over the first two Matsubara frequencies $\omega_0=\pi T$ and $\omega_1 = 3\pi T$:

\begin{equation}
    \mathrm{Im}\Sigma^{\mathrm{fit}}(i\omega_n) = \alpha_0(T) + \alpha_1(T) \omega_n,
\end{equation}

where 
\begin{align}
    \alpha_0 &= \frac{3 \mathrm{Im}\Sigma(i\omega_0)-\mathrm{Im}\Sigma(i\omega_1)}{2},\\
    \alpha_1 &= \frac{\mathrm{Im}\Sigma(i\omega_1)- \mathrm{Im}\Sigma(i\omega_0)}{2\pi T}.
\end{align}

Substituting $\mathrm{Im}\Sigma^{\mathrm{fit}}(i\omega_0)$ and $\mathrm{Im}\Sigma^{\mathrm{fit}}(i\omega_1)$ in the spectral representation \eref{eq:spectral_self}, and using the scaling function \eref{eq:scalingSupp}, we obtain the temperature dependence $T^\nu$ from a log-log fit of:
\begin{equation}
\alpha_0 \propto T^\nu.
\label{eq:alpha_0}
\end{equation}
We can also evaluate the temperature dependence $T^\nu$ with a similar approach using a quadratic fit over the first three Matsubara frequencies. In both cases, \eref{eq:alpha_0} holds, but with a different proportionality constant. The proportionality constant depends on the degree of the fit but the exponent $\nu$ does not. The results for the exponent $\nu$ are shown in \tref{tab:exposants} for both the linear fit over the two first Matsubara frequencies and the quadratic fit over the first three Matsubara frequencies.

\begin{table}
    \begin{center}
        \begin{tabular}{c c c c c}
        \hline
        \hline
          & \multicolumn{2}{c}{$x=0.20$}&\multicolumn{2}{c}{$x=0.25$}\\ 
          & Antinode &  Node & Antinode & Node  \\ \hline
          Linear fit & 1.8 & 1.4 & 1.9 &  1.8 \\
          Quadratic fit & 2.0 & 1.4 & 2.0 & 1.9  \\
        \hline \hline
        \end{tabular}
        \caption{Exponent $\nu$ for the temperature dependence $T^\nu$ of the imaginary part of the self-energy calculated assuming $\omega/T$ scaling. The linear and quadratic fit methods are detailed from \eref{eq:scalingSupp} to \eref{eq:alpha_0}.}\label{tab:exposants}
    \end{center}
\end{table}

\begin{figure}
    \centering
    \includegraphics[width=0.785\columnwidth]{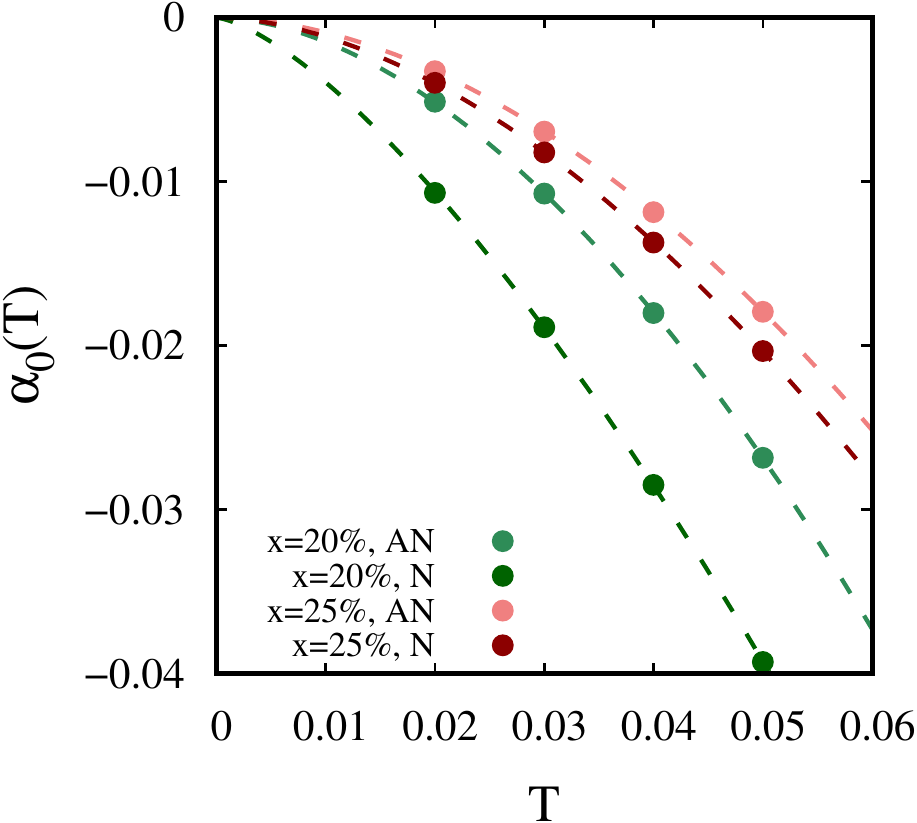} 
    \caption{Fits that gave the exponents in \tref{tab:exposants} for the linear fit method.}
    \label{fig:a0T}
\end{figure}

\figref{fig:a0T} shows the quality of the fits from \eref{eq:alpha_0} that gave the exponents in \tref{tab:exposants}, for the case where $\alpha_0$ is calculated using the linear fit over the Matsubara self-energy. 

\section{Cutoff frequency as a function of interaction strength}

Having seen that, with our chosen band parameters, the Fermi-liquid quasiparticles are well-defined at the antinode, we focus the rest of our discussion on the effects of electron-electron correlations on the Fermi liquid in this region of the Fermi surface. In a Fermi liquid, the imaginary part of the self-energy is expected to show the $\omega^2$ behaviour of \eref{eq:sigma_reel} for frequencies below a cutoff frequency $\omega_c$ \cite{Jacko_2009}. It has been proposed that the Fermi-liquid cutoff frequency can indicate the strength of the interactions in a correlated material \cite{Horio_2020}. 

To verify whether the TPSC approach can reproduce this experimental proposal, we calculate $\omega_c$ for $x=0.15$, $x=0.20$ and $x=0.25$ at the antinode for values of $U$ ranging from $1$ to $5$. The cutoff frequency is set as the frequency where the imaginary part of the Matsubara self-energy deviates from the $\omega_n^2$ behaviour of the Matsubara self-energy \eref{eq:sigma_mats}. 

\fref{fig:U} shows the cutoff frequency as a function of $U$ and as a function of $Z$. The cutoff frequency does decrease when $U$ increases. We compare these results to the measurements of $\omega_c$ in hole- and electron-doped cuprates, which show that $\omega_c$ is smaller in the hole-doped cuprate LSCO than in PLCCO \cite{Horio_2020}. This supports previous theoretical suggestions that hole-doped cuprates are more strongly correlated than electron-doped cuprates \cite{Senechal_2004, Hankevych_2006, Weber_2010}. 

It is difficult, however, to determine precisely the relation between $\omega_c$ and $U$ or $Z$. The main issue comes from the determination of $\omega_c$ itself. Here, we take $\omega_c$ as the frequency at which the relative deviation between the fit obtained from \eref{eq:SigmaMatsubara} and the calculated Matsubara self-energy reaches a threshold of $15$\%. Using a threshold of $10$\% or $20$\% for the deviation does not qualitatively change our results, as discussed in the main text. The second difficulty in the calculation of $\omega_c$ is due to the fact that we do this calculation in Matsubara frequencies. The imaginary part of the Matsubara self-energy has $\mathcal{O}(\omega_n^3)$ terms whose analytic continuation contributes to $\Sigma'$ and not to $\Sigma''$. This could lead to an underestimation of the value of $\omega_c$. Finally, both the cutoff frequency and $Z$ are momentum-dependent, making it difficult to establish a unique relation between these quantities and $U$.    
\section{Strength of interactions and a spectroscopic Kadowaki-Woods ratio}
\begin{figure}
    \centering
    \includegraphics[width=0.95\columnwidth]{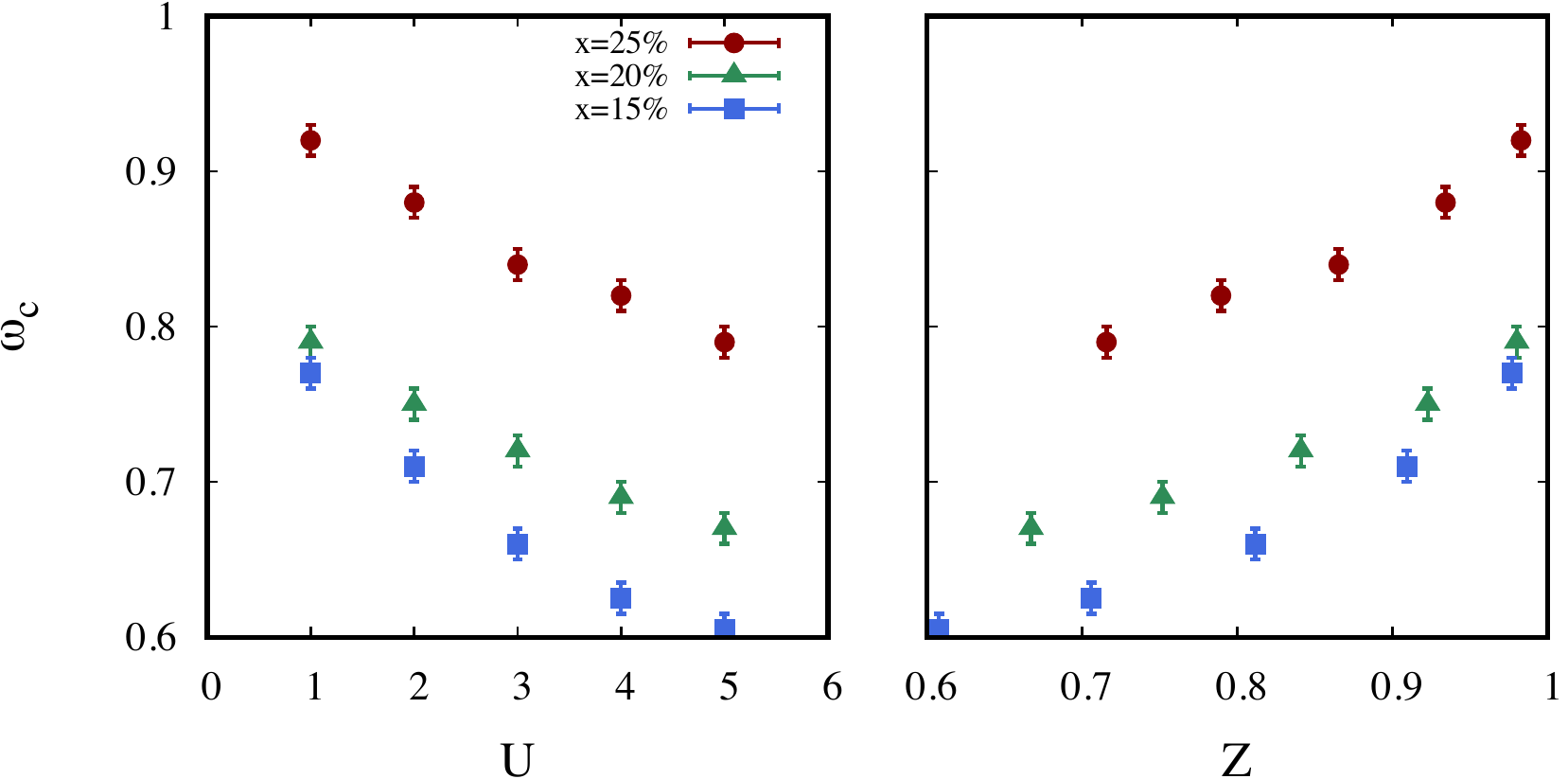} 
    \caption{Fermi-liquid cutoff frequency $\omega_c$ as a function of (left panel) Hubbard $U$ and (right panel) quasiparticle weight $Z$. All the calculations were done at the antinode, for $x=0.25$ (red), $x=0.20$ (green) and $x=0.15$ (blue). }
    \label{fig:U}
\end{figure}
In its original formulation, the Kadowaki-Woods ratio is calculated from the resistivity and the specific heat and is not actually universal. Rather, it depends on the class of materials, and can even vary within a single family of compounds \cite{Jacko_2009}. 
\begin{figure}
    \centering
    \includegraphics[width=0.8\columnwidth]{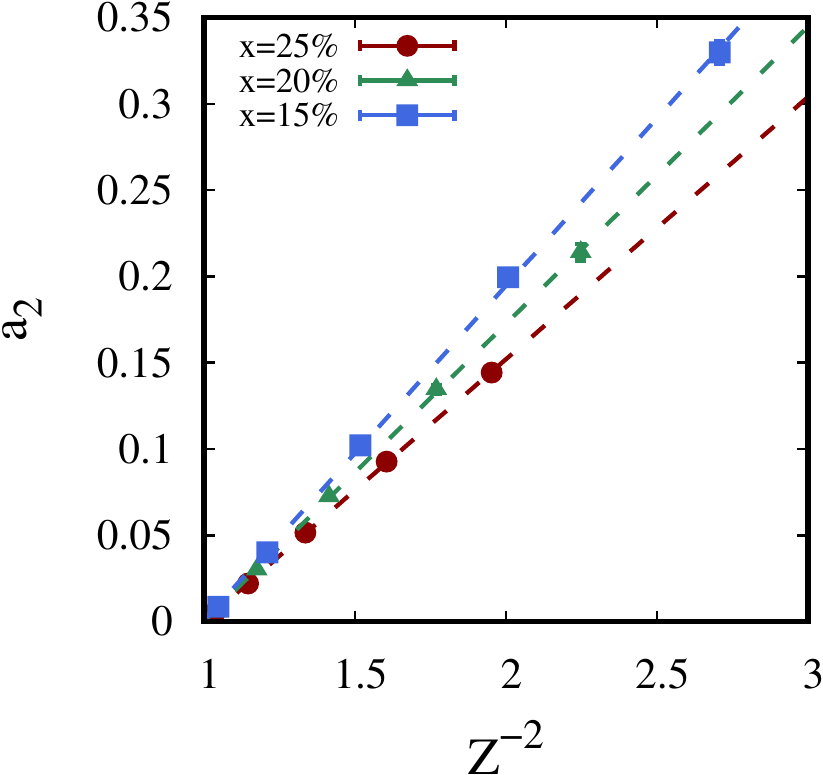} 
    \caption{Kadowaki-Woods ratio in the spectroscopic formulation introduced in Ref \cite{Horio_2020}, with $a_2$ the coefficient of the $\omega^2$ dependence $\Sigma''$ in units of $1/t$ and $Z$ the quasiparticle weight. Here, all the dots are computed at different values of $U$ (shown in \fref{fig:U}), at the antinode, for $x=0.25$ (red), $x=0.20$ (green) and $x=0.15$ (blue).}
    \label{fig:KW}
\end{figure}
We find that the spectroscopic formulation of the Kadowaki-Woods ratio $a_2\propto Z^{-2}$ proposed by Horio {\it et al.} \cite{Horio_2020} shown in \fref{fig:KW} is not universal. Even though all our calculations are performed with the same band parameters, we might speculate that the difference in dopings reflected in the density of states could explain why we do not find a universal ratio. But there is a deeper reason for the lack of universality, that we now explain. 

To investigate the meaning of the equation $a_2\propto Z^{-2}$, we use the formalism of \cite{Jacko_2009}. We write the Fermi-liquid self-energy as
\begin{align}
    \Sigma''(\omega,T) = \left\{\begin{matrix}
        -a \left ( (\pi T)^2 +\omega^2\right) &\omega<\omega_c,\\
        F(\frac{\omega}{\omega_c}) & \omega>\omega_c.
    \end{matrix} \right.
    \label{eq:selfJacko}
\end{align}
Here, $F(\omega/\omega_c)$ is a function that goes to 0 as the frequency becomes large and leaves $\Sigma''(\omega,T)$ continuous at $\omega=\omega_c$. We start by computing the self-energy in Matsubara frequencies using \eref{eq:selfJacko} and the spectral representation 
\begin{equation}
    \Sigma(i\omega_n) = \int_{-\infty}^{\infty}\frac{d\omega}{\pi}\frac{\Sigma''(\omega,T)}{\omega-i\omega_n}.
\end{equation}
From this, the previously defined coefficients $a_0$, $a_1$ and $a_2$ can be written as
\begin{align}
    a_0(T) &= -a(\pi T)^2,\\
    a_1 &= -\frac{4a\omega_c \xi}{\pi}, \label{eq:a1}\\
    a_2 &= a,
\end{align}
where $\xi$ is a constant that depends on the cutoff function $F(\omega/\omega_c)$. Following Ref. \cite{Jacko_2009}, this constant is

\begin{equation}
    \xi = \frac{1}{2}\left ( 1 - \frac{1}{a\omega_c^2}\int_1^\infty \frac{F(y)}{y^2}\right).
\end{equation}
In Ref. \cite{Jacko_2009}, Jacko \textit{et al} consider that the function $F(\omega/\omega_c)$ decreases monotonously from $\omega=\omega_c$ to $\omega\rightarrow \infty$, which means that $1/2<\xi<1$. This assumption is not valid in our calculations, and we instead find that $\xi$ ranges from $2$ to $4$ depending on the doping, the angle and the choice of the cutoff frequency.

Using a Kramers-Kronig relation to compute the real part of the retarded self-energy from the imaginary part of the retarded self-energy, \eref{eq:selfJacko} used above, yields 
\begin{equation}
    \Sigma'(\omega,T=0) = \Sigma'(\omega\rightarrow \infty,T=0) + a_1\omega_n,
\end{equation}
where the $a_1$ coefficient found from the Kramers-Kronig relation is the same as \eref{eq:a1}, which was obtained from the spectral representation for the Matsubara self-energy. This serves as another consistency check for our calculations.

 Then, from the expression for the quasiparticle weight $Z$ in \eref{eq:Za1} we obtain, expliciting the momentum dependence on the Fermi surface,

 \begin{equation}
     Z_{\mathbf{k}}^{-1}-1 = \frac{4 \xi_{\mathbf{k}}}{\pi}\omega_{c,\mathbf{k}} a_{2,\mathbf{k}} .
     \label{eq:a2vsZm1Supp}
 \end{equation}
 In \cite{Jacko_2009}, the relation $Z^{-2} \propto a_{2}$ with a proportionality coefficient that does not depend on the cutoff frequency $\omega_c$ emerges from the previous equation when we assume that: (1) Interactions are taken be strong ($Z\ll1$) so that $Z^{-1}-1 \simeq Z^{-1}$ and (2) The $a_2$ coefficient is defined as $a_2 = s/\omega_c^2$, with $s$ a material dependent constant with units of energy. In our calculations, however, neither of these assumptions is appropriate, as seen from \fref{fig:U}. Hence, we study \eref{eq:a2vsZm1Supp} without further approximation by plotting $a_2 \omega_c$ as a function of $Z^{-1}-1$ at the antinode. The results are shown in \fref{fig:a2vsZm1} in the main text. The scaling that follows from Kramers-Kronig is not expected to be universal due to the cutoff function $F(\omega/\omega_c)$, which controls the slope in \eref{eq:a2vsZm1Supp}. The cutoff function, in general, could be momentum-dependent and material-dependent.  Nevertheless, we show in the main text that this momentum and material dependence is negligible when estimating orders of magnitude.

\end{appendix}

\end{document}